\documentclass[fleqn,usenatbib]{mnras}

\usepackage[T1]{fontenc}

\DeclareRobustCommand{\VAN}[3]{#2}
\let\VANthebibliography\thebibliography
\def\thebibliography{\DeclareRobustCommand{\VAN}[3]{##3}\VANthebibliography}

\usepackage{graphicx}	
\usepackage{amsmath}	
\usepackage{amssymb}	
\usepackage{subcaption}
\usepackage{array} 
\usepackage{newtxtext,newtxmath}
\usepackage{multirow}
\usepackage{multicol}

\usepackage{xspace}
\newcommand{\mjup}{\,$M_{\rm J}$\space}
\newcommand{\mjupnospace}{\,$M_{\rm J}$}
\newcommand{\rjup}{\,$R_{\rm J}$\space}
\newcommand{\rjupnospace}{\,$R_{\rm J}$}
\newcommand{\msun}{\,$M_{\odot}$\space}
\newcommand{\msunnospace}{\,$M_{\odot}$}
\newcommand{\rsun}{\,$R_{\odot}$\space}

\newcommand{\lsun}{\,$L_{\odot}$\space}
\newcommand{\tess}{{\it TESS\space}}
\newcommand{\tessnospace}{{\it TESS}}
\newcommand{\gaia}{{\it Gaia\space}}

\newcommand{\ngts}{{NGTS\space}}
\newcommand{\ngtsnospace}{{NGTS}}
\newcommand{\feros}{{FEROS\space}}
\newcommand{\ferosnospace}{{FEROS}}
\newcommand{\LSO}{La Silla Observatory\space}
\newcommand{\PAR}{Paranal Observatory\space}
\newcommand{\teff}{{T$_{\rm eff}$\space}}

\newcommand{\vsini}{{$v\sin{i}$\space}}

\newcommand{\msininospace}{{$M\sin{i}$}}
\newcommand{\feh}{[Fe/H]\space}

\newcommand{\systemt}{{\rm TOI-2490b\space}}
\newcommand{\systemtnospace}{{\rm TOI-2490b}}
\newcommand{\systemA}{{\rm TOI-2490\space}}
\newcommand{\systemAnospace}{{\rm TOI-2490}}
\newcommand{\fsed}{$f_\mathrm{sed}$\space}
\newcommand{\fsednospace}{$f_\mathrm{sed}$}

\title[\systemt]{TOI-2490b- The most eccentric brown dwarf transiting in the brown dwarf desert}
\author[Henderson et. al.]{Beth A. Henderson$^{1}$ \thanks{E-mail: bah26@leicester.ac.uk}, 
Sarah L. Casewell$^{1}$,
Andr\'es Jord\'an$^{2,3,4}$,
Rafael Brahm$^{2,3,4}$,
Thomas Henning$^{5}$,
\newauthor Samuel Gill$^{6,7}$,
L.C. Mayorga$^{8}$,
Carl Ziegler$^{9}$,
Keivan G.\ Stassun$^{10}$
Michael R. Goad$^{1}$,
Jack Acton $^{1}$,
\newauthor  Douglas~R.~Alves $^{11}$,
David~R.~Anderson$^{12}$,
 Ioannis Apergis $^{6,7}$,
David~J.~Armstrong$^{6,7}$,
Daniel Bayliss $^{6,7}$,
\newauthor  Matthew~R.~Burleigh$^{1}$,
Diana Dragomir$^{13}$,
 Edward Gillen$^{14}$,
 Maximilian N. Günther$^{15}$,
 Christina Hedges$^{16}$,
\newauthor Katharine~M.~Hesse$^{17}$,
Melissa J. Hobson$^{2,5}$,
James S. Jenkins$^{18,19}$,
Jon M. Jenkins$^{20}$,
Alicia Kendall$^{1}$,
\newauthor Monika Lendl$^{21}$,
 Michael B. Lund$^{22}$,
James McCormac$^{6,7}$,
 Maximiliano~Moyano$^{11}$,
Ares Osborn$^{6,7}$,
\newauthor Marcelo Tala Pinto$^{2,3}$,
 Gavin Ramsay$^{23}$,
David Rapetti$^{20, 24}$,
 Suman Saha$^{16,18}$,
 Sara Seager$^{13,25,26}$, 
\newauthor Trifon Trifonov$^{5,27,28}$,
 Stéphane Udry$^{21}$,
 Jose~I.~Vines$^{11}$,
Richard~G.~West $^{6,7}$,
 Peter~J.~Wheatley $^{6,7}$,
\newauthor Joshua~N.~Winn$^{29}$,
Tafadzwa Zivave$^{6,7}$\\
$^{1}$ School of Physics and Astronomy, University of Leicester, University Road, Leicester LE1 7RH, UK\\
$^{2}$Facultad de Ingenier\'ia y Ciencias, Universidad Adolfo Ib\'a\~nez, Av.\ Diagonal las Torres 2640, Pe\~nalol\'en, Santiago, Chile\\
$^{3}$Millennium Institute of Astrophysics, Santiago, Chile.\\
$^{4}$Data Observatory Foundation, Chile\\
$^{5}$Max-Planck-Institut f\"ur Astronomie, K\"onigstuhl 17, Heidelberg 69117, Germany\\
$^{6}$Centre for Exoplanets and Habitability, University of Warwick, Gibbet Hill Road, Coventry, CV4 7AL, UK\\
$^{7}$Dept. of Physics, University of Warwick, Gibbet Hill Road, Coventry, CV4 7AL, UK\\
$^{8}$ The Johns Hopkins University Applied Physics Laboratory, 11100 Johns Hopkins Rd, Laurel, MD, 20723, USA\\
$^{9}$ Department of Physics, Engineering and Astronomy, Stephen F. Austin State University, 1936 North St, Nacogdoches, TX 75962, USA\\
$^{10}$ Department of Physics and Astronomy, Vanderbilt University, Nashville, TN 37235, USA\\
$^{11}$ Departamento de Astronom\'ia, Universidad de Chile, Casilla 36-D, Santiago, Chile\\
$^{12}$ Instituto de Astronom\'ia, Universidad Cat\'olica del Norte, Angamos 0610, 1270709, Antofagasta, Chile\\
$^{13}$ Department of Physics and Astronomy, University of New Mexico, 210 Yale Blvd NE, Albuquerque, NM 87106, USA\\
$^{14}$ Astronomy Unit, Queen Mary University of London, Mile End Road, London E1 4NS, UK\\
$^{15}$ European Space Agency (ESA), European Space Research and Technology Centre (ESTEC), Keplerlaan 1, 2201 AZ Noordwijk, The Netherlands\\
$^{16}$ NASA Goddard Space Flight Center, 8800 Greenbelt Rd, Greenbelt, MD 20771, USA\\
$^{17}$ Department of Physics and Kavli Institute for Astrophysics and Space Research, Massachusetts Institute of Technology, Cambridge, MA 02139, USA\\
$^{18}$ Instituto de Estudios Astrofísicos, Universidad Diego Portales,  Av. Ej\'ercito 441, Santiago, Chile\\
$^{19}$Centro de Astrof\'isica y Tecnolog\'ias Afines (CATA), Casilla 36-D, Santiago, Chile\\
$^{20}$ NASA Ames Research Center, Moffett Field, CA 94035, USA\\
$^{21}$ Observatoire de Gen{\`e}ve, Universit{\'e} de Gen{\`e}ve, Chemin Pegasi 51, 1290 Versoix, Switzerland\\
$^{22}$NASA Exoplanet Science Institute, IPAC, California Institute of Technology, Pasadena, CA 91125 USA\\
$^{23}$ Armagh Observatory and Planetarium, College Hill, Armagh, BT61 9DG, UK\\
$^{24}$Research Institute for Advanced Computer Science, Universities Space Research Association, Washington, DC 20024, USA\\
$^{25}$ Department of Earth, Atmospheric and Planetary Sciences, Massachusetts Institute of Technology, Cambridge, MA 02139, USA\\
$^{26}$ Department of Aeronautics and Astronautics, MIT, 77 Massachusetts Avenue, Cambridge, MA 02139, USA \\
$^{27}$ Department of Astronomy, Sofia University ``St Kliment Ohridski'', 5 James Bourchier Blvd, BG-1164 Sofia, Bulgaria\\
$^{28}$ Landessternwarte, Zentrum f\"ur Astronomie der Universt\"at Heidelberg, K\"onigstuhl 12, 69117 Heidelberg, Germany\\
$^{29}$ Department of Astrophysical Sciences, Princeton University, Princeton, NJ 08544, USA\\
}

\date{Accepted XXX. Received YYY; in original form ZZZ}

\pubyear{2024}

\begin{document}
\label{firstpage}
\pagerange{\pageref{firstpage}--\pageref{lastpage}}
\maketitle

\begin{abstract}
We report the discovery of the most eccentric transiting brown dwarf  in the brown dwarf desert, \systemtnospace. The brown dwarf desert is the lack of brown dwarfs around main sequence stars within $\sim3$~AU and is thought to be caused by differences in formation mechanisms between a star and planet. To date, only $\sim40$ transiting brown dwarfs have been confirmed. \systemt is a $73.6\pm2.4$ \mjupnospace, $1.00\pm0.02$ \rjup brown dwarf orbiting a $1.004_{-0.022}^{+0.031}$ \msunnospace, $1.105_{-0.012}^{+0.012}$ \rsun sun-like star on a 60.33~d orbit with an eccentricity of $0.77989\pm0.00049$. The discovery was detected within \tess sectors 5 (30 minute cadence) and 32 (2 minute and 20 second cadence). It was then confirmed with 31 radial velocity measurements with \feros by the WINE collaboration and  photometric observations with the Next Generation Transit Survey. Stellar modelling of the host star estimates an age of $\sim8$~Gyr, which is supported by estimations from kinematics likely placing the object within the thin disc. However, this is not consistent with model brown dwarf isochrones for the system age suggesting an inflated radius. Only one other transiting brown dwarf with an eccentricity higher than 0.6 is currently known in the brown dwarf desert. Demographic studies of brown dwarfs have suggested such high eccentricity is indicative of stellar formation mechanisms. 

\end{abstract}




\section{Introduction}

Brown dwarfs are sub-stellar objects which canonically fall within the 13-80 \mjup mass range (\citealp{burgasser,baraffe02,burrows93}). They are not quite massive enough to burn hydrogen in their cores, but some can burn deuterium (\citealp{whitworth18,bate02}). As they age, they cool, progressing through the M-L-T-Y spectral types, and their radii contract. The first confirmed brown dwarfs were Gliese 229B \citep{gliese} and Teide 1 \citep{teide} discovered in 1995, although before this, brown dwarf candidates were found using the radial velocity method \citep{rvbd}. Since then, thousands of brown dwarfs have been discovered, most of which do not have any known companions \citep{ultracool}.

Using radial velocities to find brown dwarf candidates orbiting main sequence stars has been very successful (e.g. \citealt{hatp13c}), and has led to the discovery of eccentric brown dwarfs \citep{HD191760b,HD91669B,HD137510}. However, with the radial velocity technique, the true companion mass cannot be determined; instead, we learn only the companion's minimum possible mass (\msininospace). Unless the brown dwarf also transits its host star, we have no model independent measure of the radius. If the radius can be determined, then the mass-radius-age degeneracy can be broken and an age estimate can be found using model isochrones (\citealp{baraffe03,baraffe15,marley21}). 

To determine the radii of objects in binary systems we use the transit method. This has been extremely successful for planets, with thousands confirmed to date. Due to brown dwarfs having radii which are similar in size to Jupiter, they should be as easy to detect in transiting exoplanet surveys as Hot Jupiters. However, there are very few brown dwarfs that have been discovered transiting within 3 AU of main sequence stars. The rarity of such objects has been called the `brown dwarf desert' (e.g. \citealt{grether06}) and their frequency of occurrence is at a local minimum between those of two other populations -- planets and low mass stars. The minimum is thought to be caused by differences in formation mechanisms between planets and stars. Giant planets are thought to form via core accretion, producing masses up to a few Jupiter masses \citep{mordasini09} whereas stars in binaries are thought to form via disc fragmentation \citep{bate02}.

To date, only $\sim40$ transiting brown dwarfs have been found in the `desert'. Surveys have tried to fill this parameter space \citep{triaud17} but have been unable to. Many have also looked into this population of objects to determine any trends due to properties such as eccentricity, mass or metallicity. For example, from an analysis of all known brown dwarf companions within 3 AU of a solar-type star, \citet{mage14} proposed that they formed a split population. The lower-mass population (up to 42.5 \mjup) exhibited an eccentricity distribution in which the mean eccentricity decreases with increasing mass. The higher-mass population however, which lies above 42.5 \mjupnospace, was not found to have any dependence of mean eccentricity upon mass. This led \citet{mage14} to conclude that the lower mass brown dwarfs may have formed like planets, whereas the higher mass brown dwarf population formed more like stars. Other works have attempted to confirm this split population \citep{grieves21} but have been unable to. Due to this, some have questioned if there are indeed two distinct populations \citep{carmichael19}.

Other properties of these transiting brown dwarfs have also been discussed in the literature, but eccentricity is considered to be a clear indicator of formation mechanism for brown dwarfs, as long as the eccentricity has not been affected by other external factors. For instance effects relating to tidal interactions \citep{jackson08} can circularise the orbit, eradicating any initial eccentricity that would otherwise have been a clue about the formation mechanism. If the host star is in a binary, then the outer companion can also affect the eccentricity of the inner companion's orbit due to Kozai-Lidov perturbations \citep{kozailidov}. 

\citet{bowler20} used directly imaged giant planets (9 objects, 2 - 15~\mjup) and brown dwarfs (18 objects, 15 - 75~\mjup) to explore eccentricity distributions on a population level. Their sample avoids external effects on the eccentricity of a system, due to tidal forces and Kozai-Lidov perturbations, as much as possible by choosing objects which have wide orbits (between 5-100 AU) and not choosing objects where the host star is part of a known binary. Some objects do have wider, tertiary companions that could affect their inner companion over longer timescales, but this effect is unlikely to be large enough to have affected the inner companion.  \citet{bowler20} found that objects with higher mass-ratios tend to have a wide spread of eccentricities, compared with low mass-ratio objects.  \citet{bowler20} suggest that this points to high mass-ratio objects having formed via stellar formation mechanisms. They also found a suggestion of two populations when separating based on mass (2-15 \mjup vs 15-75 \mjup) indicating that brown dwarfs at wider separations are more likely to have formed via stellar formation mechanisms. 

In this paper, we report the discovery of \systemtnospace, a new, highly eccentric transiting brown dwarf discovered using \tessnospace.  This system is one of few close, transiting systems that is not circularised, and as such is an excellent test of the connection between formation and eccentricity. We first discuss the observations of \systemAnospace, the host star to our newly discovered brown dwarf in Section \ref{sec:obs}. We then discuss how stellar modelling of \systemA was done and how we used its parameters to model \systemt in Section \ref{sec:analysis}. In Section \ref{sec:discussion}, we discuss where \systemt lies within the population of brown dwarf desert objects, estimate the age of the system using kinematics, and explore how the temperature of the atmosphere of \systemt could be affected by its extreme orbit around its host star. 

\section{Observations}\label{sec:obs}

\begin{figure}
    \centering
    \includegraphics[width=\linewidth, trim =12cm 8cm 12cm 8cm, clip]{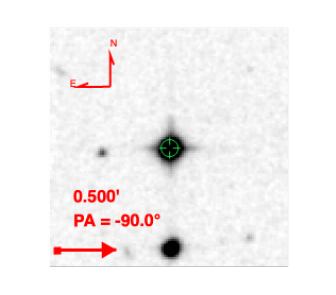}
    \caption{DSS image showing \systemA (green) with other nearby objects. North and East are shown on the figure in the top left corner as well as a 0.5~arcmin position angle (PA) arrow for scale.}
    \label{fig:dss}
\end{figure}

\systemA was first discovered using a systematic search of single transit events within the Transiting Exoplanet Survey Satellite (\tessnospace, \citealt{tess}) lightcurves, which were extracted from the full-frame images \citep{spoc, caldwell20}. The orbital period was then constrained using multiple cameras from the Next Generation Transit Survey (\ngtsnospace, \citealt{NGTS}). In addition to the photometry, 31 radial velocity data points were obtained using \feros \citep{feros}. Each of the photometric observations are listed in Table \ref{tab:phot} and the radial velocity observations are shown in Table \ref{tab:radial_velocities}. Figure \ref{fig:dss} shows the Deep Sky Survey (DSS) image of \systemAnospace.

\begin{table*}
\caption{Photometric observations made for \systemAnospace, with details about the bandpass, cadence, observation dates and number of observed transits for each instrument.}              
\label{tab:phot}      
\begin{tabular}{l c c c c }          
\hline\hline                        
Instrument & Bandpass & Cadence & Number  & Observation \\
&&&of transits&date\\
\hline 
\ngts & 520-890 nm & 13 s & 1 & 2022 Nov 20\\
\tess Sector 32& 600-1000 nm & 20 s & 1 & 2020 Nov 19 to 2020 Dec 17\\
\tess Sector 5& 600-1000 nm & 1800 s & 1 & 2018 Nov 15 to 2018 Dec 11\\
\hline
\end{tabular}
\end{table*}

\subsection{TESS}\label{sec:tess}

\begin{figure*}
    \centering
    \includegraphics[width=\linewidth]{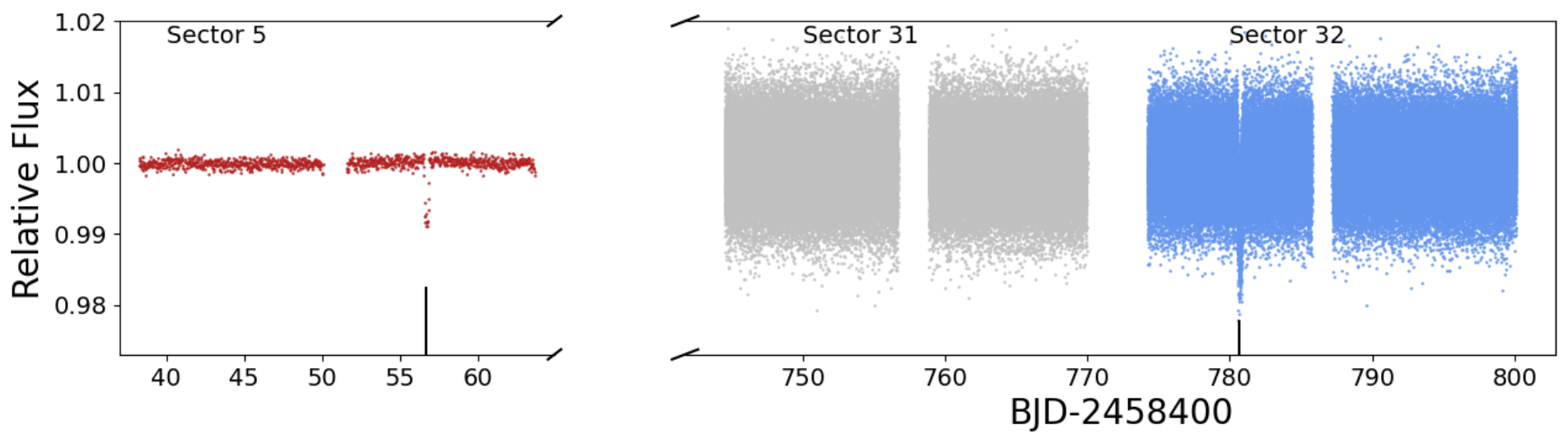}
    \caption{\tess timeseries lightcurve of \systemAnospace, showing the relative flux of the 30-minute (1800 s) Sector 5 (red) and 20-second cadence Sector 31 (grey) and 32 (blue) data.}
    \label{fig:tess}
\end{figure*}

\tess obtained 30 minute cadence photometry of \systemAnospace\, (TIC 77437543) in Sector 5 which was observed from 2018 November 15 to 2018 December 11. It also obtained both 2-minute and 20-second cadence photometry within Sector 32, observing from 2020 November 19 to 2020 December 17. Within this sector, one transit was observed. \systemA was also observed in Sector 31, but no transits were visible, as seen in Figure \ref{fig:tess}.

The $TESS$ Sector 31 and 32 data were processed by the Science Processing Operations Center (SPOC – \citealt{spoc}) at NASA Ames Research Center and the Full Frame Image (FFI) data from Sector 5 were processed by the SPOC as part of the TESS-SPOC project \citep{caldwell20}.

 All data were downloaded from and are publicly available in the Mikulski Archive for Space Telescopes\footnote{https://mast.stsci.edu}. We use the PDCSAP lightcurve which removes systematics in the data (e.g. \citealt{Stumpe2012, smith12, Stumpe2014}). We normalised the data using the out-of-transit data. The systematic search for single transit events is described in full within \citet{mono}. 

\begin{figure}
    \centering
    \includegraphics[width=\linewidth]{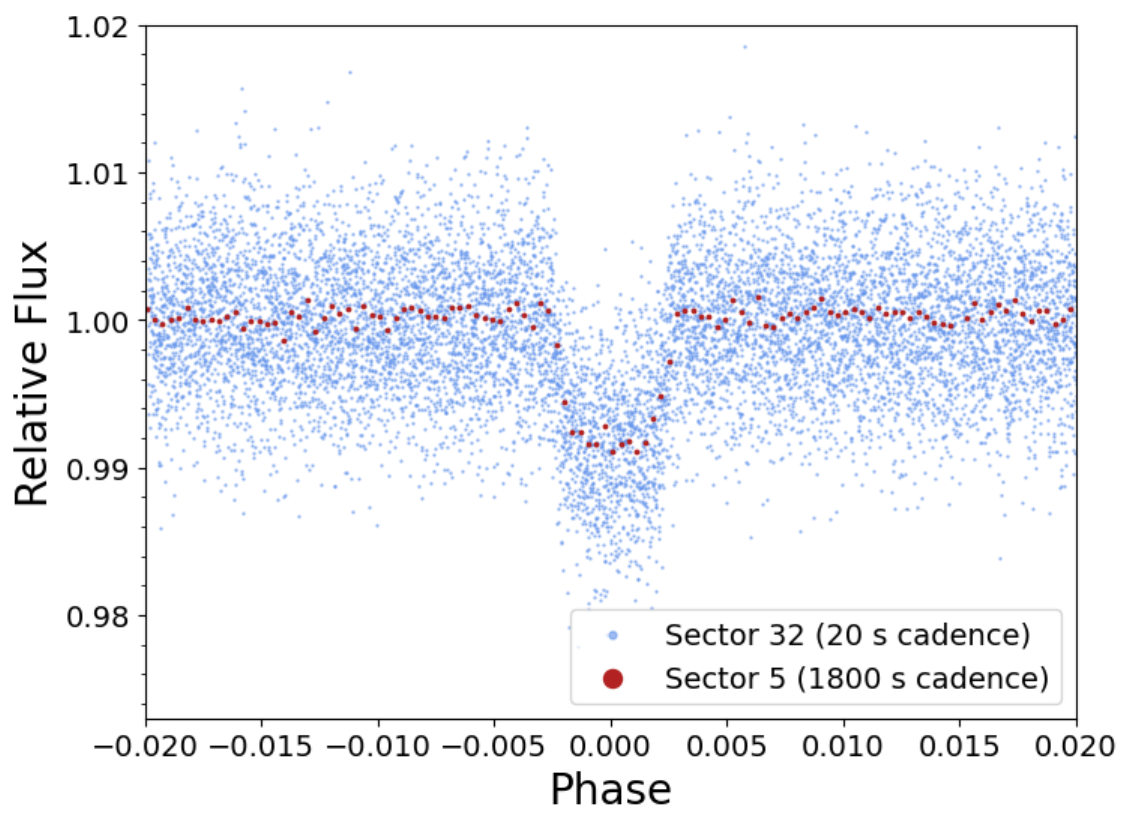}
    \caption{\tess phase-folded lightcurve of \systemAnospace, showing the relative flux of the 30-minute (1800 s) Sector 5 (red) and 20-second cadence 32 (blue) data, zoomed in on the transit.}
    \label{fig:tess_phase}
\end{figure}

The SPOC detected the single transit of TOI-2490b in Sector 32 and a multi-sector search of Sectors 31 and 32 using an adaptive matched filter \citep{jenkins02, jenkins20}. The transits occurred in Sector~5 at BJD~2458456.699 and Sector~32 at BJD~2459180.691, with a detected depth of 10~ppt and a total transit duration of $\sim7.5$~h in both sectors (Figure \ref{fig:tess_phase}). While the orbital period attributed to TOI-2490 b was incorrect, the transit signature passed all the diagnostic tests conducted and reported in the Data Validation reports \citep{Twicken} but for the odd/even transit depth test (as expected), and the location of the target star was localised to within 0.728$\pm$2.5” of the transit source location in the Sector 32 search. This difference image centroiding test is important in constraining the likelihood of background eclipsing binary contamination as it typically constrains the transit source location to a region much smaller than a pixel, as is the case here.  As can be seen from Figure \ref{fig:tess}, \systemA also shows no sign of magnetic activity, such as flares. We also checked the \tess lightcurves for potential false positives.

 Once we were confident that the transit was real, we proceeded to make further observations to constrain the period and ephemeris. From the two transits in \tessnospace, we were able to limit the possibilities for the period to be $P=724$/N days where N is an integer number of orbits. The Sector 31 data constrained this further, due to no detected transit.

While PDCSAP light curves are corrected for crowding when they are generated,  there are sometimes unknown errors in the input catalogue used to predict the crowding, and the Point Spread Function of the instrument does change over time, which is not captured in the static Pixel Response Functions used. \tess data  contains CROWDSAP values for each object, and this value indicates how much flux within the aperture is estimated to be from the target source, with the rest coming from contaminating sources. A value of 1 would therefore be completely uncontaminated. There is minimal dilution expected within \tessnospace, supported by the CROWDSAP value of 0.956 for Sector 5 and 0.971 for Sector 32. While the closest object is $\sim34$ arcsec away (\gaia $G$=18.06), with another bright object $\sim50$ arcsec away (\gaia $G$=13.68), the target is brighter (\gaia $G$=11.76), but we expect some dilution due to the large pixel scale of \tess ($\sim 21$ arcsec). Therefore we fit for dilution of the \tess data in the global modelling of \systemtnospace.

\subsection{FEROS}\label{sec:feros}

\begin{table*}
\caption{Radial velocity measurements and the associated uncertainties from \feros for \systemAnospace. The Bisector span (BIS), BIS error, signal-to-noise (SNR) and exposure time are also listed.}              
\label{tab:radial_velocities}      
\centering   
\begin{tabular}{l c c c c c c}          
\hline\hline                        
BJD & Radial velocity [$\rm km\, \rm s^{-1}$] & Radial velocity error [$\rm km\, \rm s^{-1}$] & BIS [$\rm km\, \rm s^{-1}$] & BIS error [$\rm km\, \rm s^{-1}$] & SNR & Exposure time [s]\\
\hline 
2458911.56481&24.8321&0.0062&-0.016&0.01&91.0&1200.0\\
2458916.54839&24.516&0.0118&-0.032&0.017&41.0&1200.0\\
2458923.52455&23.8884&0.0071&-0.009&0.011&76.0&1200.0\\
2459188.64668&24.9431&0.0064&-0.02&0.01&86.0&1200.0\\
2459190.6115&25.4725&0.006&-0.026&0.01&94.0&1200.0\\
2459191.62828&25.5773&0.0065&-0.027&0.01&85.0&1200.0\\
2459194.64213&25.6286&0.0078&-0.005&0.012&67.0&1200.1\\
2459196.58966&25.5944&0.0059&-0.02&0.01&96.0&1200.0\\
2459206.62583&25.1183&0.0079&-0.021&0.012&65.0&1200.0\\
2459209.63082&24.9711&0.0088&-0.036&0.013&58.0&1200.0\\
2459211.63541&24.8363&0.0065&-0.017&0.01&85.0&1200.0\\
2459213.69205&24.7354&0.0077&-0.014&0.012&68.0&1200.0\\
2459219.6793&24.2879&0.0066&-0.019&0.011&82.0&1200.0\\
2459260.59879&25.4327&0.0081&-0.004&0.012&64.0&1200.0\\
2459264.62326&25.2548&0.0067&-0.021&0.011&82.0&1200.0\\
2459272.58322&24.7773&0.0066&-0.022&0.011&83.0&1200.0\\
2459281.57743&24.1333&0.007&-0.003&0.011&77.0&1200.0\\
2459482.76511&19.4139&0.0067&-0.024&0.011&80.0&1200.0\\
2459485.78238&14.614&0.0084&-0.029&0.013&61.0&1200.0\\
2459493.79588&25.4996&0.0075&-0.044&0.012&70.0&1800.0\\
2459497.78606&25.5209&0.0081&-0.054&0.012&64.0&1200.0\\
2459500.80887&25.4001&0.0081&-0.083&0.012&64.0&1200.0\\
2459506.75271&25.1431&0.0061&-0.031&0.01&91.0&1200.0\\
2459650.5977&23.2974&0.0077&-0.027&0.012&68.0&1200.0\\
2459653.58076&22.871&0.0077&-0.052&0.012&68.0&1200.1\\
2459656.57936&22.388&0.0089&-0.015&0.013&58.0&1200.0\\
2459679.52716&25.4096&0.0085&-0.004&0.013&61.0&1500.0\\
2459684.50581&25.2317&0.0087&-0.042&0.013&59.0&1200.0\\
2459687.48247&25.0795&0.0072&-0.007&0.011&74.0&1200.0\\
2459688.48546&24.9657&0.0088&-0.057&0.013&58.0&1200.0\\
2459704.46458&23.9409&0.0078&-0.024&0.012&67.0&1200.0\\
\hline
\end{tabular}
\end{table*}

To model the mass of \systemtnospace, we obtained radial velocity data of \systemA with \ferosnospace. \feros is an echelle spectrograph, based at \LSO on the 2.2~m telescope \citep{feros}. \feros works in the 360 to 920 nm wavelength range, with a resolution of 48000 \citep{feros}. \feros also utilises two fibres, one on target and one that can either be on the sky background or the Thorium-Argon-Neon calibration lamp \citep{feros}. Our data used the simultaneous lamp observation method. There is a 2.7~arcsecs aperture, and the data reduction follows the \texttt{ceres} automated pipeline \citep{ceres}. \texttt{ceres} performs the CCD reduction and chooses the optimal extraction for the spectra \citep{ceres}. It also wavelength calibrates, performs an instrumental drift correction and normalises the spectra, computing radial velocities and bisector span measurements using the cross-correlation technique with a G2 mask \citep{ceres}.

\begin{figure*}
    \centering
    \subfloat[]{\includegraphics[]{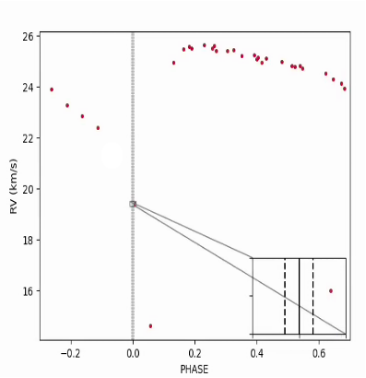}}
    \qquad
    \subfloat[]{\includegraphics[scale=0.58]{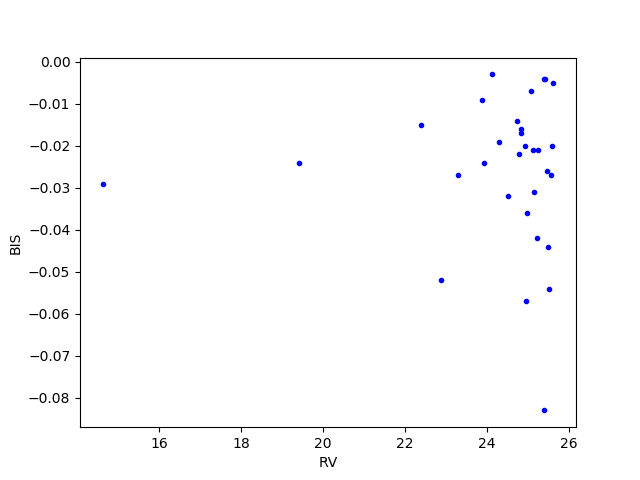}}
    \caption{\textbf{(a)}: Phase-folded radial velocity points from \ferosnospace. An inset around the transit is provided to show the point does not fall within the transit window. The solid vertical line shows phase 0 (centre of transit) and the dashed vertical lines show the estimated edges of the transit, based on the transit width. \textbf{(b)}: BIS are plotted with the radial velocity points and show no trend that would indicate a background object or an additional object in the system.}
    \label{fig:rvphase}
\end{figure*}

Table \ref{tab:radial_velocities} shows the set of 31 radial velocity measurements taken, including their respective errors, bisector span (BIS), signal-to-noise (SNR) and exposure time. The data used 20 minute exposures, except for one point which had a 25 minute exposure. We obtained good signal-to-noise ratios (SNRs) for all the data points (ranging from 41.0 to 96.0) and obtained a mean radial velocity error of 8 $\rm m\, \rm s^{-1}$. The radial velocity data provided confirmation of an orbital period of 60.33 days, one of the potential periods found with the \ngts transit. The BIS showed no obvious trend with the data, as shown in Figure \ref{fig:rvphase}. 

Figure \ref{fig:rvphase} shows the phase-folded radial velocity points from Table \ref{tab:radial_velocities} using the found period of 60.33~d. From this figure, a high level of eccentricity can be seen. We have also plotted a zoomed in view of the transit window to show that the point close to 0 phase does not fall within the transit.

\subsection{NGTS}\label{sec:ngts}

Multi-camera follow up observations of \systemt were taken by \ngtsnospace, a twelve-telescope array situated at ESO's \PAR, Chile \citep{NGTS}. The twelve, 20~cm  telescopes cover around 100 deg$^2$ of sky and operate in a custom bandpass of 520 nm to 890 nm \citep{NGTS}. The high precision photometry provided by \ngts was ideal for following up the transit we observed within the \tess data. More information about the \ngts facility is provided in \citet{NGTS}.

We started observing the target, in order to constrain potential orbital periods, on 2021 November 04, processing the images following  the pipeline in \citet{NGTS}. We observed the transit successfully on 2022 November 20, with between 2232 and 2254 science images with a 13~s cadence, on four cameras simultaneously. Once extracted, the fluxes from each camera were co-added, normalised and binned to two minutes. The two-minute binned data was used in the global modelling of this object. 

\begin{figure}
    \centering
    \includegraphics[width=\linewidth]{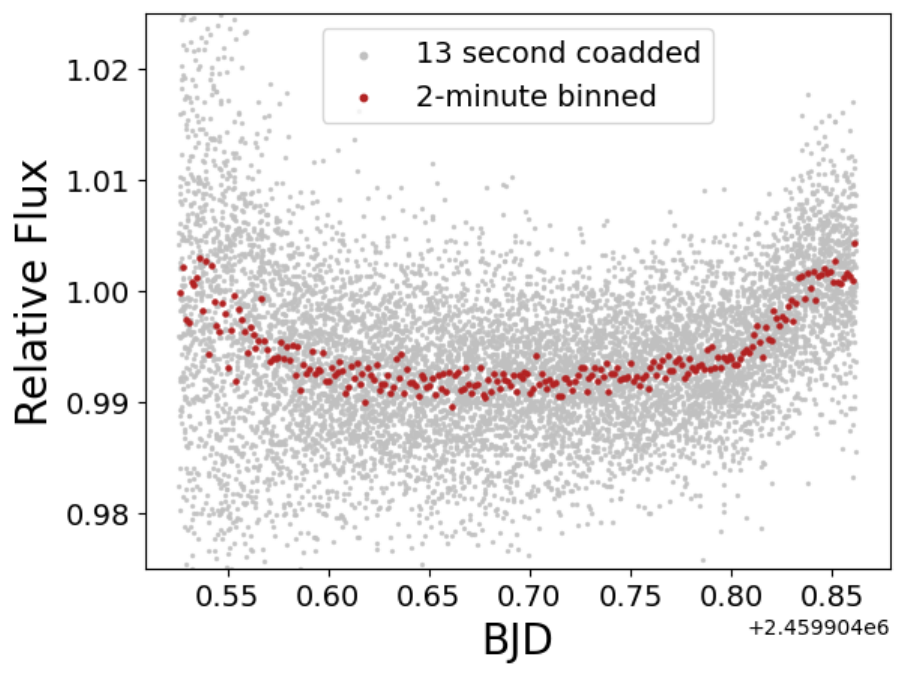}
    \caption{\ngts lightcurve of \systemAnospace, showing the relative flux of the 13-second, co-added data from four cameras (grey) and the two-minute binned, co-added data (red).}
    \label{fig:ngtslc}
\end{figure}

The lightcurve of the two-minute, co-added data can be seen in Figure \ref{fig:ngtslc}. The transit occurred at around BJD~2459904.68 with a transit depth of $\sim9$~ppt and total transit duration of $\sim7.4$~h. As can be seen in Figure \ref{fig:ngtslc}, we just about observed the whole transit, with some of out-of-transit data. This is due to the long total transit duration, which covers the entire observability window for \ngtsnospace. The two-minute binned data had a transit depth which matched well with the 20-second cadence data from \tessnospace, once normalised. This means we obtained enough out-of-transit data to normalise the lightcurve and provide an accurate transit depth.

Figure \ref{fig:dss} shows the DSS image of \systemA and other nearby objects. As the closest object is much fainter (\gaia G=18.06) and $\sim 34$ arcsec away, we can safely assume there is no dilution within the 15~arcsec radius aperture, with an \ngts pixel scale of $\sim5$ arcsec.

\section{Analysis}\label{sec:analysis}

\subsection{Nearby objects}\label{sec:cpmc}

\begin{figure}
    \centering
    \includegraphics[width=\linewidth]{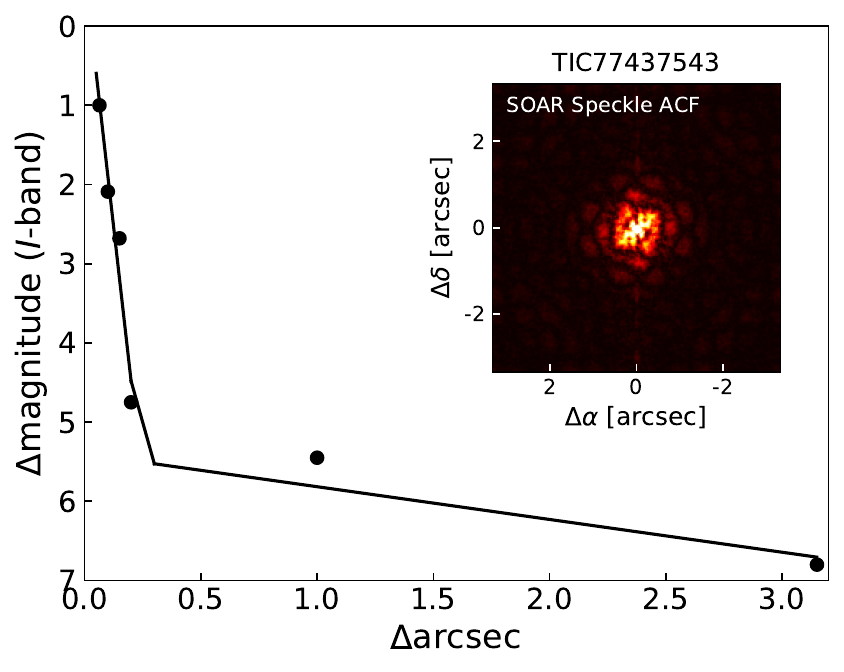}
    \caption{SOAR Speckle imaging of \systemA in the $I$ band showing the 5$\sigma$ detection sensitivity and auto-correlation function. No nearby stars were detected.}
    \label{fig:speckle}
\end{figure}

High-angular resolution imaging is needed to search for nearby sources that can contaminate the \tess photometry, resulting in an underestimated radius of the occulting object, or be the source of astrophysical false positives, such as background eclipsing binaries. We searched for stellar companions to \systemA with speckle imaging on the 4.1-m Southern Astrophysical Research (SOAR) telescope \citep{Tokovinin} on 27 February 2021 UT, observing in Cousins $I$ band, a similar visible bandpass to \tess. This observation was sensitive to a 5.5-magnitude fainter star at an angular distance of 1 arcsec from the target. More details of the observations within the SOAR TESS survey are available in \citet{ziegler}. The 5$\sigma$ detection sensitivity and speckle auto-correlation functions from the observations are shown in Figure \ref{fig:speckle}. No nearby stars were detected within 3\arcsec of TOI-2490 in the SOAR observations.

Within the 15~arcsec \ngts aperture, there are no sources in the \gaia DR3 catalogue \citep{gaiadr3}, other than the target. However, within 60~arcsec there are two other objects. One is much fainter (\gaia G=18.06) and does not have parallax or proper motion values which are similar to \systemAnospace. The other object (\gaia ID 4818818482795434112 and TIC77437539) is a 13.68 \gaia G-magnitude object with a parallax value $3.71\pm0.01$ which is consistent within $3\sigma$ of \systemAnospace, and a radial velocity value $24.65\pm1.22$ $\rm km\, \rm s^{-1}$ which is consistent within $1\sigma$ of \systemAnospace. The $\mu$RA ($19.70\pm0.01$ $\rm mas\, \rm yr^{-1}$) and $\mu$DEC ($12.36\pm0.01$ $\rm mas\, \rm yr^{-1}$) values are $\sim0.25$ $\rm mas\, \rm yr^{-1}$ difference which is much greater than a $5\sigma$ difference. There are no other objects with similar $\mu$RA and $\mu$Dec values within 2~arcmin of \systemAnospace. \systemA and TIC77437539 are separated by $\sim50$~arcsec on the sky, a projected separation of $\sim13382$~AU. Therefore, even if TIC77437539 is a common proper motion companion, it will not effect the orbital radial velocity measurements observed by \ferosnospace. As stated in Section \ref{sec:tess}, these objects will slightly dilute the \tess photometry and are accounted for within the global modelling of \systemtnospace.

\subsection{TOI-2490: Host star parameters}\label{sec:host}

\begin{figure}
    \centering
    \includegraphics[width=\linewidth]{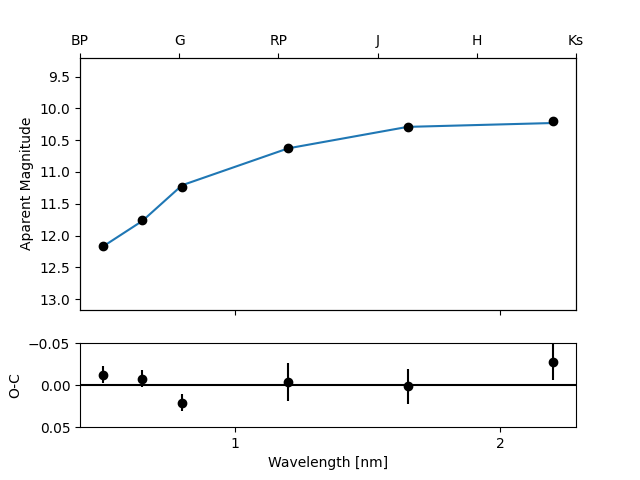}
    \caption{Results of the SED fit of \systemA, using photometry listed in Table \ref{tab:systemAparameters}.}
    \label{fig:sed}
\end{figure}

We performed a detailed analysis of the host star to determine its main physical properties. We followed the iterative procedure presented in \citet{hd1397}. Briefly, we obtain the atmospheric stellar parameters ($T_{\rm eff}$, $\log{g}$, [Fe/H],$v\sin{i}$) from the co-added \feros spectra using the \texttt{zaspe} package \citep{zaspe}, that identifies the best-fitted synthetic spectrum generated with the ATLAS9 stellar models. We then perform a spectral energy distribution (SED) fit using publicly available broad band photometry, the Gaia DR3 parallax,  the \texttt{zaspe} derived $T_{\rm eff}$ as a prior, and we assume a metallicity fixed to the value found by \texttt{zaspe}(Figure \ref{fig:sed}). The SED fitting produces a more precise estimate of $\log{g}$, which is then used in a new iteration of \texttt{zaspe}. We continue with iterations until reaching convergence in $\log{g}$. We find that \systemA is a solar-like G-type main sequence star, but with a significant metal enrichment ([Fe/H] = $+0.32 \pm 0.05$) and a likely age of 7.9$_{-1.8}^{+1.5}$~Gyr.

Table \ref{tab:systemAparameters} gives the proper motion, parallax and magnitudes from \gaia DR3 \citep{gaia,gaiadr3,gaiaextra}, 2MASS \citep{2mass}, \tess from ExoFOP \citep{tess} and SDSS  \citep{sdsscam,sdssphot,sdsstech}. 

\begin{table}
\caption{Magnitudes, parameters and kinematics of \systemA are from \gaia DR3 \citep{gaia,gaiadr3,gaiaextra}, 2MASS \citep{2mass}, \tess from ExoFOP \citep{tess} and SDSS  \citep{sdsscam,sdssphot,sdsstech}. \systemAnospace's fitted parameters were found using the method discussed in Section \ref{sec:host}.}           
\label{tab:systemAparameters}      
\centering   
%

\begin{tabular}{l l}          
\hline\hline                        
Parameter & \systemA \\
\hline 
\gaia Source ID & 4818818585874648320\\
RA  & $73.1245128086102$\\
Dec  & $-36.25709623870886$\\

\\
$\mu$RA [$\rm mas\, \rm yr^{-1}$] & $19.45\pm 0.01$ \\
$\mu$Dec [$\rm mas\, \rm yr^{-1}$] & $12.61 \pm 0.02$\\
\gaia Radial Velocity [$\rm km\, \rm s^{-1}$]& $24.63\pm1.00$\\
Parallax [$\rm mas$] & $3.74\pm 0.01$\\
Distance [pc] & $267.65\pm0.97$ \\ \\

\textbf{Magnitudes}\\
\gaia $G$ & $11.7562^{+0.0002}_{-0.0001}$ \\
\gaia $BP$ & $12.1332^{+0.0007}_{-0.0008}$\\
\gaia $RP$ & $11.2175\pm0.0003$\\
\tess [$T$]  & $11.2759\pm0.006$\\
SDSS$_{g}$ & $12.2580\pm0.0330$\\ 
SDSS$_{r}$ & $11.7170\pm0.0060$\\ 
SDSS$_{i}$ & $11.9030\pm0.5620$\\ 
2MASS$_{J}$ & $10.6250\pm0.0230$\\
2MASS$_{H}$ & $10.2930\pm0.0210$\\
2MASS$_{Ks}$ & $10.2030\pm0.0210$\\ \\

\textbf{Fitted Parameters}\\
$\rm T_{\rm eff}$ $\rm(K)$ &$5558.0\pm80$\\
$\log g$ (dex)  & $4.353_{-0.017}^{+0.020}$\\
$\rm [Fe/H]$ (dex) & $0.32\pm0.05$ \\
$M_{\rm A}$ [$M_{\odot}$] &$1.004_{-0.022}^{+0.031}$\\
$R_{\rm A}$ [$R_{\odot}$] & $1.105\pm0.012$\\
Age [Gyr] &$7.9_{-1.8}^{+1.5}$\\
\vsini [$\rm km\, \rm s^{-1}$] & $2.8\pm0.3$\\
Av [mag] & $0.073_{-0.049}^{+0.068}$\\
L [\lsun] & $1.053_{-0.033}^{+0.048}$\\
$\rho$ [g/cm$^3$] & $1.049_{-0.051}^{+0.059}$\\
\hline
\multicolumn{2}{l}{}
\end{tabular}
\end{table}

\subsection{TOI-2490b Parameters}\label{sec:fitted}

Global modelling for \systemt was performed using \textsc{Allesfitter} (\citealt{allesfitter-code,allesfitter-paper}). \textsc{Allesfitter} fits photometry and radial velocity data simultaneously to produce estimations for parameters of the system (\citealt{allesfitter-code, allesfitter-paper}). \textsc{Allesfitter} makes use of models such as \textsc{ellc} \citep{ellc} and samplers such as \textsc{Dynesty} \citep{dynesty} and \textsc{emcee} \citep{emcee}. For modelling \systemtnospace, we used dynamic nested sampling methods to estimate the best fitting parameters to the data, with 500 live points.

\begin{figure*}
    \centering
    \subfloat[]{\includegraphics[scale=0.48]{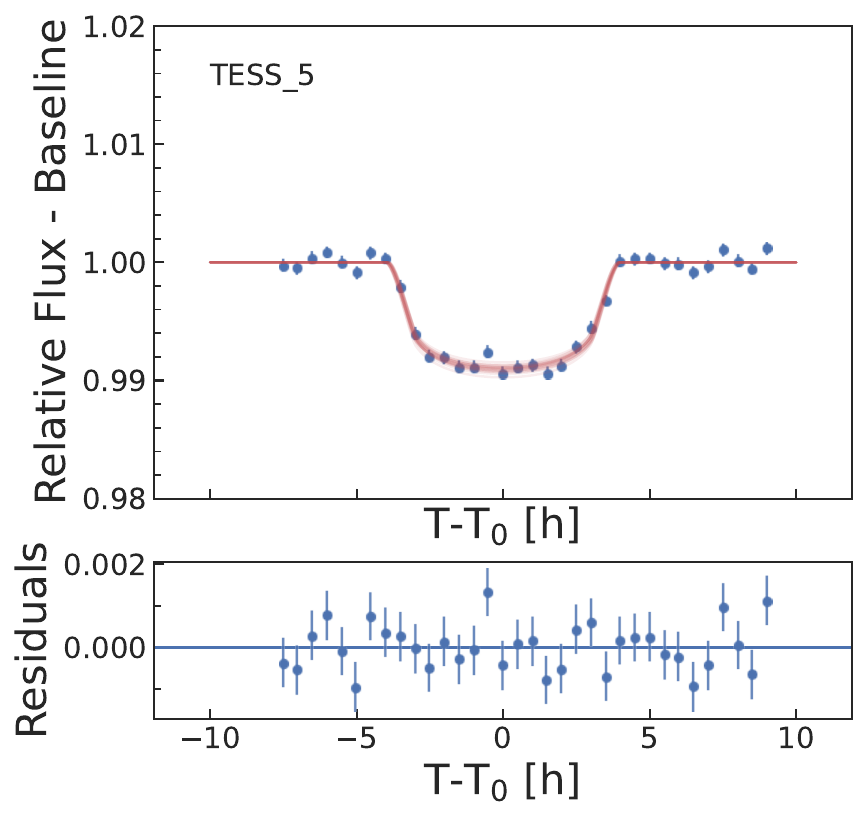}}
    \qquad
    \subfloat[]{\includegraphics[scale=0.48]{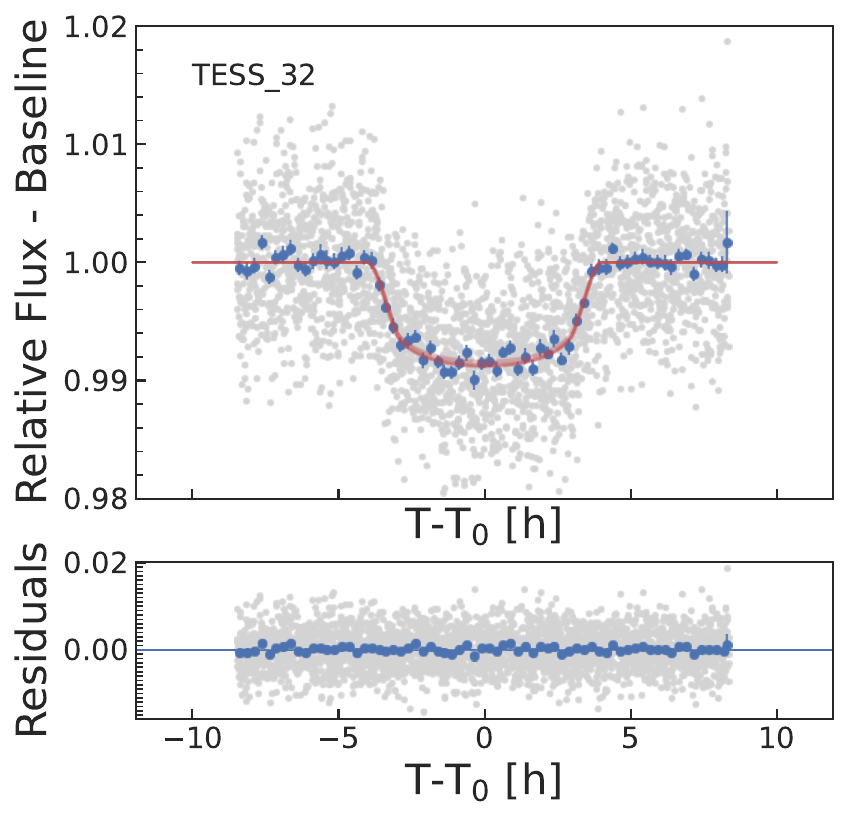}}
    \qquad
    \subfloat[]{\includegraphics[scale=0.48]{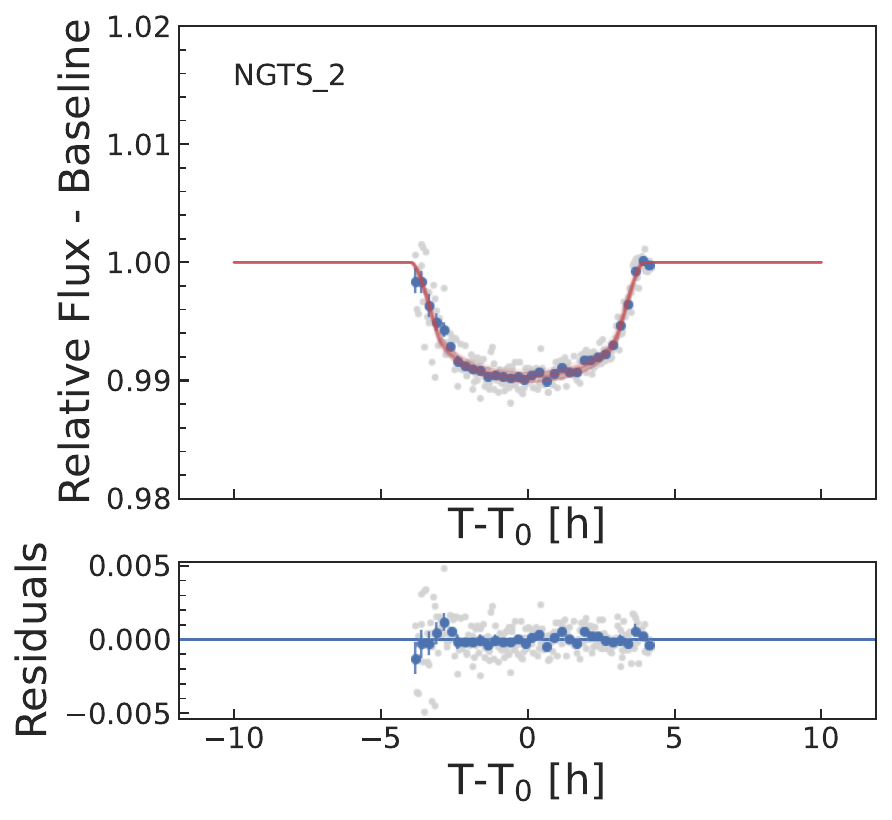}}
    \qquad
    \subfloat[]{\includegraphics[scale=0.48]{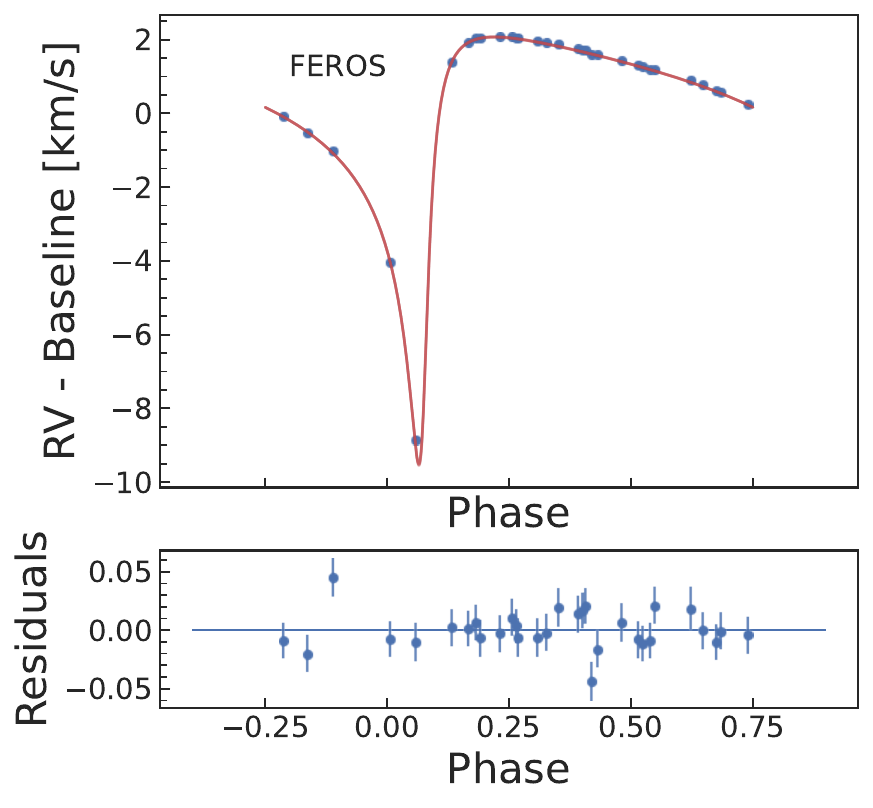}}
    \caption{Model fits (in red) to: \textbf{(a)} \tess 30 minute data in Sector 5, \textbf{(b)} \tess 20~s data in Sector 32, \textbf{(c)} \ngts data (binned to 2 minutes) and \textbf{(d)} \feros data from \textsc{allesfitter}. The red lines show 20 samples drawn from the posterior to the fit with \textsc{Allesfitter}. All plots share the same axes scale for ease of comparison.}
    \label{fig:model_fit}
\end{figure*}

For the priors to the fit, we used a combination of normally distributed and uniformly distributed parameter spaces. Due to the set up of \textsc{Allesfitter}, all parameters must be given an initial prior, as well as a parameter space. Any parameters with initial priors based on the data have been fit with uniform parameter spaces, to avoid biasing the result. All fitted parameters, their prior values and parameter spaces can be seen in Table \ref{tab:parameters} and are discussed within this section. 

We first estimated limb darkening constants for \ngts and \tess with the Python Limb Darkening ToolKit (\textsc{ldtk}, \citet{ldtk}), using the Kipping triangular quadratic method \citep{kipping13}. These were then used as priors, with a parameters space normally-distributed around the prior value, using the errors from \textsc{ldtk} as the standard deviation.

We fixed the dilution of \ngts to 0 as no known objects in \gaia are within the aperture for \ngtsnospace. We allowed the dilution values for both the \tess sector data sets to vary uniformly between 0 and 1 with an initial value of 0. \textsc{Allesfitter} uses the radius ratio and a scaled sum of radii ($(R_\star + R_b) / a_b$) as a prior for the fit. To calculate the initial value for the radius ratio we used a rough estimate of the transit depth from the \ngts data and fit with a uniform and wide distribution. For $(R_\star + R_b) / a_b$ we used the radii from \systemAnospace, estimated the radii of \systemt using our prior for the radius ratio and a rough estimate of semi-major axis from Kepler's third law. We also fit $(R_\star + R_b) / a_b$ with a uniform and wide parameter space.

For the eccentricity parameters $\sqrt{e_b} \cos{\omega_b}$ and $\sqrt{e_b} \sin{\omega_b}$, as well as the cosine of the inclination, we used wide uniform parameter spaces and started at 0. We fit the period with a uniform parameter space, using the period of 60.33 d solved by \feros as the initial prior. We used half of the range of our radial velocity values as our prior for semi-amplitude, with a wide and uniform parameter space. The epoch was fit with a uniform parameter space around a rough estimate of the epoch from \tess Sector 32 but was allowed to shift to avoid correlation between the period and Epoch.

The logarithm of the white noise was modelled per data set, with jitter being modelled for \ferosnospace. The parameter space for all datasets was uniform between 0 and -10, with the initial prior being an estimate based on the mean of the relative flux or radial velocity errors.

We used hybrid offset baselines for all photometry data-sets as we could see the baselines were flat. A hybrid offset baseline uses the median of the residuals to offset the data at each sampling step \citep{allesfitter-code,allesfitter-paper}. We chose to use hybrid offsets for the photometry instead of sample offsets (where a constant value is subtracted from the data) because the sample offsets over-estimated the baseline and ignored the transit.

We tested multiple baselines for the \feros radial velocities, making use of the $\ln{Z}$ tool in \textsc{allesfitter} to choose the best fit to the data. $\ln{Z}$ is the logarithmic Bayesian evidence, with
\begin{equation}
    Z := P(D \mid M) ,
\end{equation}
where $Z$ is the probability ($P$) of obtaining the data ($D$) given a certain model ($M$), as detailed in \citet{allesfitter-paper}. In nested sampling, when the change in this value is less than a certain threshold (0.01 in our case) it is considered to be converged \citep{allesfitter-paper}. This means, when a singular parameter at a time is changed (in this case the \feros baseline) the $\ln{Z}$ value can be compared using the logarithmic Bayes Factor ($\ln{R}$),

\begin{equation}
    \ln{R} = \ln{Z_{2}} - \ln{Z_{1}} ,
\end{equation}
where, $\ln{R}$ is the difference between the logarithmic Bayesian evidence of model 2 ($\ln{Z_{2}}$) and model 1 ($\ln{Z_{1}}$). If $\ln{R}$ is less than 0, model 1 is preferred. We use the interpretation of the Bayes factor from \citet{jeffreys98theory} and \citet{kass} listed in \citet{allesfitter-paper}, where $\ln{R}$ greater than 4.6 is 'decisive' and model 2 should be preferred.

We tested sample offset, hybrid offset and hybrid spline baselines for the \feros data. A hybrid spline baseline fits the data with a smooth spline at each sampling step \citep{allesfitter-code,allesfitter-paper}. The $\ln{Z}$ values were $13151.11 \pm 0.38$, $13160.75 \pm 0.79$ and $13199.61 \pm 0.32$ for sample offset, hybrid offset and hybrid spline respectively. We chose the hybrid spline as our preferred baseline for the \feros data due to $\ln{R}$ being greater than 5.

The modelled data and the fitted residuals are shown in Figure \ref{fig:model_fit}. The resultant values for the fitted and derived parameters can be seen in Table \ref{tab:parameters}, with the derived corner plot in Figure \ref{fig:derived_corner}. Full versions of these tables can be seen in Section \ref{appendix:tables}. We determined that \systemt has a mass of $73.6\pm2.4$ \mjupnospace, a radius of $1.00\pm0.02$ \rjupnospace, an eccentricity of $0.77989\pm0.00049$ and a period of 60.33 d. This means \systemt is a long-period, high-mass brown dwarf with the largest eccentricity of all known transiting brown dwarfs within the brown dwarf desert.

\begin{table*}
\caption{The fitted and derived parameters for \systemtnospace, produced by \textsc{Allesfitter}.  More detailed parameters are in tables \ref{tab:systemAparameters} and \ref{tab:derived_parameters_full}.}           
\label{tab:parameters}      
\centering   
\begin{tabular}{l c c c } 
\hline\hline
parameter & symbol & prior, parameter space & value\\ 
\hline
\multicolumn{4}{c}{\textbf{Fitted parameters}} \\ 
\textbf{Limb Darkening Parameters}\\
\ngts LDC 1 & $q_{1;\mathrm{NGTS}}$ & 0.6218, $\mathcal{N}(0.6218, 0.0032)$ & $0.6133\pm0.0028$ \\ 
\ngts LDC 2 & $q_{2;\mathrm{NGTS}}$ & 0.2397, $\mathcal{N}(0.2397, 0.0007)$ & $0.2398\pm0.0006$ \\ 
\tess Sector 32 LDC 1 & $q_{1;\mathrm{TS32}}$ & 0.5067, $\mathcal{N}(0.5067, 0.0024)$ & $0.5070\pm0.0023$ \\ 
\tess Sector 32 LDC 2 & $q_{2;\mathrm{TS32}}$ & 0.2221, $\mathcal{N}(0.2221, 0.0006)$ & $0.2221\pm0.0005$ \\ 
\tess Sector 5 LDC 1 & $q_{1;\mathrm{TS5}}$ & 0.5067, $\mathcal{N}(0.5067, 0.0024)$ & $0.5069\pm0.0023$ \\ 
\tess Sector 5 LDC 2 & $q_{2;\mathrm{TS5}}$ & 0.2221, $\mathcal{N}(0.2221, 0.0006)$  & $0.2221\pm0.0005$ \\ 
\\
\textbf{System Parameters}\\
Radius ratio & $R_b / R_\star$ & 0.095, $\mathcal{U}(0.0, 0.3)$ & $0.0929\pm0.0012$ \\ 
Scaled summed radius &$(R_\star + R_b) / a_b$ & 0.019,$\mathcal{U}(0.0, 0.3)$ & $0.01822\pm0.00022$ \\ 
Cosine inclination & $\cos{i_b}$ & 0, $\mathcal{U}(0.0, 0.5)$ & $0.0140\pm0.0005$ \\ 
Epoch ($\mathrm{BJD}$) & $T_{0;b}$ & 2459180.69, $\mathcal{U}(2459175, 2459185)$ & $2459180.69430$\\
&&&$\pm0.00079$\\ 
Orbital Period (d) & $P_{orb}$ & 60.3326, $\mathcal{U}(60, 61)$ & $60.33266\pm0.00008$\\ 
RV Semi-amplitude ($\mathrm{km/s})$ & $K_b$ & 5.5, $\mathcal{U}(4, 8)$ & $5.791\pm0.011$\\ 
$\sqrt{e_b} \cos{\omega_b}$ & $f_{c}$& 0, $\mathcal{U}(-1, 1)$ & $-0.7280\pm0.0007$ \\ 
$\sqrt{e_b} \sin{\omega_b}$ & $f_{s}$& 0, $\mathcal{U}(-1, 1)$ & $-0.5000\pm0.0012$ \\ 
\\
\textbf{Dilution values}\\
\ngts Dilution & $D_{\mathrm{NGTS}}$ & - & $0.0$ (fixed) \\ 
\tess Sector 32 Dilution & $D_{\mathrm{TS32}}$ & 0, $\mathcal{U}(0, 1)$ & $0.092\pm0.029$ \\ 
\tess Sector 5 Dilution & $D_{\mathrm{TS5}}$ & 0, $\mathcal{U}(0,1)$ & $0.068_{-0.031}^{+0.032}$ \\ 
\\
\textbf{White noise}\\
- &$\log{\sigma_\mathrm{NGTS}}$ & -5.088158303984515, $\mathcal{U}(-10,0)$ &$-6.828\pm0.046$ \\ 
- &$\log{\sigma_\mathrm{TESS S32}}$ & -5.443456144580072, $\mathcal{U}(-10, 0)$ & $-5.429\pm0.013$ \\ 
- &$\log{\sigma_\mathrm{TESS S5}}$ & -7.557485934938554, $\mathcal{U}(-10, 0)$ & $-7.52_{-0.12}^{+0.13}$ \\ 
-  ($\mathrm{km/s}$) &$\log{\sigma_\mathrm{jitter}} (RV_\mathrm{FEROS})$ & -4.887276434429315, $\mathcal{U}(-10, 0)$ &$-4.30\pm0.19$\\ 
\\
\multicolumn{4}{c}{\textbf{Derived parameters}} \\ 
\systemt radius (\rjup) & $R_\mathrm{b}$ & - & $0.999\pm0.017$ \\ 
semi-major axis (AU) & $a_\mathrm{b}$ & - & $0.308\pm0.005$ \\ 
- & $R_\mathrm{A}/a_\mathrm{b}$ & - & $0.0167\pm0.0002$\\ 
- & $R_\mathrm{b}/a_\mathrm{b}$ & - & $0.00155\pm0.00003$\\ 
Inclination (deg) & $i_\mathrm{b}$ & - & $89.20\pm0.03$\\ 
Eccentricity & $e_\mathrm{b}$ & - & $0.7799\pm0.0005$\\ 
Argument of periastron (deg) & $w_\mathrm{b}$ & - & $214.48\pm0.09$\\ 
Mass ratio & $q_\mathrm{b}$ & - & $0.070\pm0.001$ \\ 
\systemt Mass (\mjup) & $M_\mathrm{b}$ & - & $73.6\pm2.4$ \\ 
Impact parameter & $b_\mathrm{tra;b}$ & - & $0.590\pm0.015$ \\ 
Total transit duration (h) & $T_\mathrm{tot;b}$ & - & $7.925\pm0.054$\\ 
Equilibrium Temperature (K) & $T_\mathrm{eq;b}$ & - & $464.2\pm7.2$\\ 

\hline
\end{tabular}
\end{table*}

\section{Discussion}\label{sec:discussion}

\subsection{Mass-radius relation}\label{sec:massrad}

\begin{figure*}
    \centering
    \includegraphics[width=\linewidth]{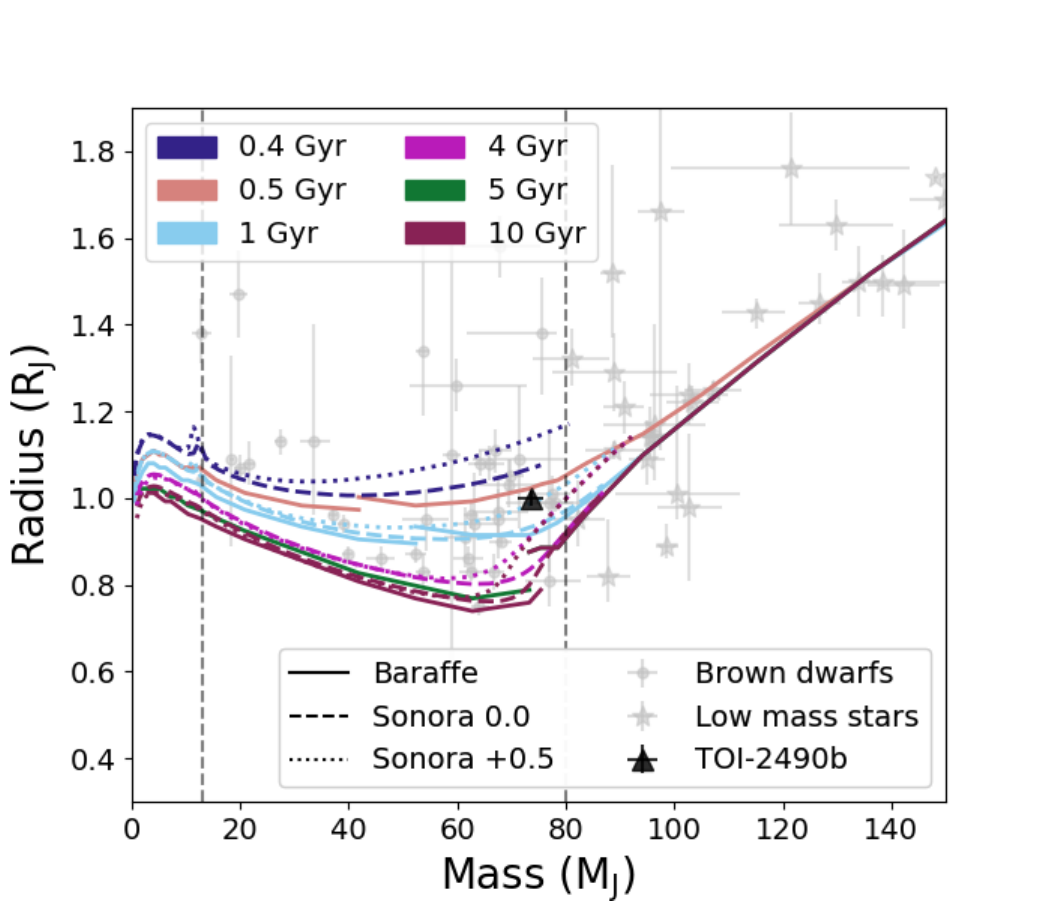}
    \caption{The mass-radius plot for all known transiting brown dwarfs and low mass stars from 12 to 150 \mjupnospace, excluding RIK-72b \citep{rik72b} and 2MASS J05352184–0546085 from \citet{Stassun06} due to their youth and therefore large radii. Grey circles are brown dwarfs, whereas grey stars are low mass stars. The dashed vertical lines highlight the 13 and 80 \mjup borders for the brown dwarf regime. \systemt is the black triangle. Model isochrones are also plotted: solid lines are the \citet{baraffe03,baraffe15} iscochrones for 0.5 (pink), 1 (light blue), 5 (green) and 10 (dark mauve) Gyr. The breaks in these isochrones are where we switch to the updated \citet{baraffe15} iscochrones from \citet{baraffe03}. The dashed and dotted isochrones show the \citet{marley21} Sonora Bobcat models for \feh of 0.0 dex and +0.5 dex respectively. These are plotted for 0.4 (dark blue), 1 (light blue), 4 (bright purple) and 10 (dark mauve) Gyr. All objects can be seen in Table \ref{appendix:tables} along with their sources. This table has been updated from \citet{grieves21} and \citet{henderson24}. }
    \label{fig:massrad}
\end{figure*}

Figure \ref{fig:massrad} shows the mass-radius plot for all known transiting objects between 10 and 150 \mjupnospace. We have included an updated table from \citet{grieves21} and \citet{henderson24} in Table \ref{appendix:tables} with information of all the objects in Figure \ref{fig:massrad}. Some objects have radii which have been updated from \citet{imporvedradii}. From this plot, we can see that \systemt falls within the brown dwarf regime, near the hydrogen burning limit.

We also plotted the \citet{baraffe03} and \citet{baraffe15} isochrones (solid lines) for substellar objects. The discontinuity in the isochrones shows where the 2003 isochrones end and the 2015 isochrones start, with the lower mass values being from the 2003 models. We also plot the \citet{marley21} Sonora Bobcat substellar isochrones for \feh 0.0 dex (dashed lines) and +0.5 dex (dotted lines). Brown dwarfs are degenerate objects, and their radii decrease over time as they cool, so young objects are expected to have larger radii than older objects, although after $\sim$4~Gyr, the radii are all very similar. Using the model isochrones for the brown dwarf (\systemtnospace), our measured mass and radius give an estimated age of 1 Gyr using the \citet{marley21} +0.5 dex isochrone. We would expect a larger radius than for a solar metallicity field brown dwarf, due to \systemAnospace's high metallicity  ($0.32\pm0.05$), however, 1~Gyr is significantly younger than was determined by SED fitting to the stellar spectrum.  We explore other ways to estimate the age of the system in Section \ref{sec:age}, and discuss the possibility that the brown dwarf's radius is inflated, giving a much younger age estimation, in Section \ref{sec:rad}. 

\subsection{Eccentricity}\label{sec:ecc}

\begin{figure}
    \centering
    \includegraphics[width=\linewidth]{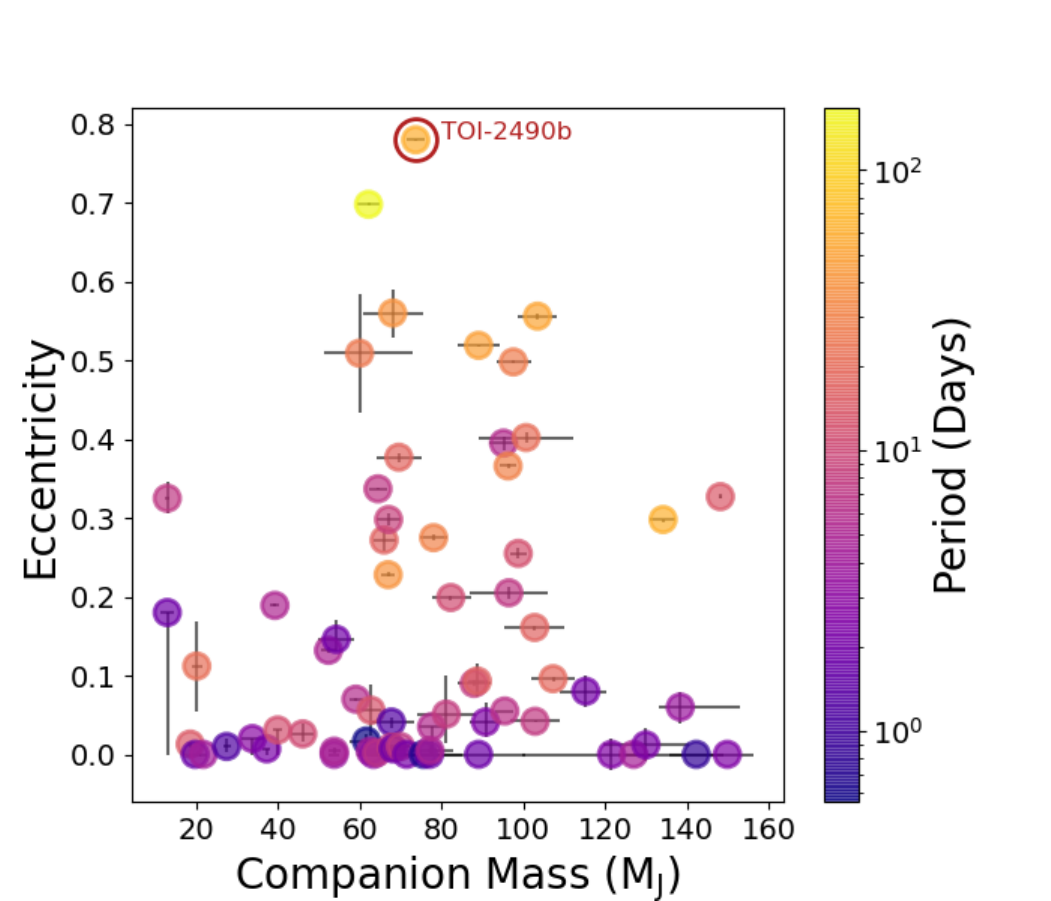}
    \caption{Mass-eccentricity plot for all objects in Table \ref{tab:full_obj_lists}. The colour indicates the orbital period. \systemt is highlighted with a red circle. RIK-72b \citep{rik72b} and 2MASS J05352184–0546085 from \citet{Stassun06} are excluded due to their youth.}
    \label{fig:massecc}
\end{figure}

We derived an eccentricity for \systemt of $0.77989\pm0.00049$. We used the method from \citet{Lucy71} to determine the significance of this detection ($>$ 2.45$\sigma_{e}$). We find that our eccentricity is hundreds of times greater the standard error, and we can state that the eccentricity of \systemt is the highest of any transiting brown dwarf in the desert. This can also be seen in Figure \ref{fig:massecc}.

We tested if \systemt should have circularised within its lifetime using tidal circularisation timescale calculations from \citet{jackson08}. Tidal interactions within a system cause the orbit of the companion to circularise over time, discussed in \citet{zahn89}. An estimate for this timescale can be calculated using equations \eqref{eq:circstar}, \eqref{eq:circb} and \eqref{eq:circe}, which are from \citet{jackson08}:

\begin{equation}\label{eq:circstar}
   \frac{1}{\tau_{circ,\star}} =  \frac{171}{16} \sqrt{\frac{G}{M_{\star}}}
   \frac{R_{\star}^{5} M_{b}}{Q_{\star}} a^{\frac{-13}{2}} ,
\end{equation}
\begin{equation}\label{eq:circb}
    \frac{1}{\tau_{circ,b}} =  \frac{63}{4} \frac{\sqrt{G M_{\star}^{3}} R_{b}^{5}}{Q_{b} M_{b}} a^{\frac{-13}{2}} ,
\end{equation}
\begin{equation}\label{eq:circe}
    \frac{1}{\tau_{e}} =  \frac{1}{\tau_{circ,\star}} + \frac{1}{\tau_{circ,b}} ,
\end{equation}
where, $\tau_{circ,\star}$, $\tau_{circ,b}$ and $\tau_{e}$ are the circularisation timescales for the star, companion and system respectively. $G$ is the gravitational constant, $M_{\star}$ and $M_{b}$ are the host and companion masses respectively, $a$ is the orbital radius and $Q_{\star}$ and $Q_{b}$ are the tidal quality factors. This will estimate a circularisation timescale with assumptions such as the orbit is not tidally locked, having a period less than 10 d and the eccentricity being low \citep{jackson08}. 

We use our results from the global modelling to estimate the circularisation timescale for \systemt for a range of $Q_{\star}$ and $Q_{b}$ values. We follow a similar method from \citet{carmichael20} and \citet{19b}. We vary the tidal quality factors as they are not well constrained in the literature, although there are values for the 3~Gyr old CWW~89Ab of $Q_{\star} > 10^9$ and $Q_{b} > 10^{4.15}$ \citep{beatty18} and $Q_{b} > 10^{4.5}$ for  2MASS J05352184–0546085 \citep{heller10}. The results can be seen in Table \ref{tab:circ_time}, where we explored a similar parameter space to that in \citet{beatty18}. From Table \ref{tab:circ_time}, it is clear that the timescale of circularisation for this system is extremely long ($10^{11}$ - $10^{14}$ yrs). From this, we are able to conclude that the high eccentricity is not unexpected, and indicates stellar formation mechanisms as in \citet{bowler20}. 

\begin{table}
\caption{Estimated tidal circularisation timescales for \systemtnospace. These values show a range of possible timescales for $Q_{s}$ values from $10^{4.5}$ to $10^{7}$ and $Q_{b}$ values from $10^{4.5}$ to $10^{6}$.}              
\label{tab:circ_time}      
\centering   
\begin{tabular}{l l c} 
\hline\hline  
$Q_{\star}$ & $Q_{b}$ & $\tau_{e}$ [Yrs] \\
\hline
\multirow{4}{*}{$10^{4.5}$}& $10^{4.5}$ & $8.85 \times 10^{11}$\\
&$10^{5}$&$8.86 \times 10^{11}$\\
&$10^{5.5}$&$8.87 \times 10^{11}$\\
&$10^{6}$&$8.87 \times 10^{11}$\\
\hline
\multirow{4}{*}{$10^{5}$}& $10^{4.5}$ & $2.79 \times 10^{12}$\\
&$10^{5}$&$2.80 \times 10^{12}$\\
&$10^{5.5}$&$2.80 \times 10^{12}$\\
&$10^{6}$&$2.80 \times 10^{12}$\\
\hline
\multirow{4}{*}{$10^{5.5}$}& $10^{4.5}$ &$8.71 \times 10^{12}$\\
&$10^{5}$&$8.82 \times 10^{12}$\\
&$10^{5.5}$&$8.85 \times 10^{12}$\\
&$10^{6}$&$8.86 \times 10^{12}$\\
\hline
\multirow{4}{*}{$10^{6}$}& $10^{4.5}$ &$2.65 \times 10^{13}$\\
&$10^{5}$&$2.75 \times 10^{13}$\\
&$10^{5.5}$&$2.79 \times 10^{13}$\\
&$10^{6}$&$2.80 \times 10^{13}$\\
\hline
\multirow{4}{*}{$10^{7}$}& $10^{4.5}$ &$1.77 \times 10^{14}$\\
&$10^{5}$&$2.37 \times 10^{14}$\\
&$10^{5.5}$&$2.65 \times 10^{14}$\\
&$10^{6}$&$2.75 \times 10^{14}$\\
\hline
\end{tabular}
\end{table}

We investigated whether the nearby object (TIC77437539) could be a common proper motion companion, and if so, could its presence have affected the orbit of \systemt due to Kozai-Lidov perturbations \citep{kozailidov}. We calculated the Kozai-Lidov timescale and found that even with an eccentricity of 0.99999 (assuming its current separation as the semi-major axis) the timescale is still many times longer than the Hubble time. If instead we assume a more realistic eccentricity of 0.59 from \citet{binaryavecc}, we achieve an even larger timescale for Kozai-Lidov perturbations. It is therefore likely that the high eccentricity of \systemt is likely due to its formation.

\subsection{System age}\label{sec:age}

Since our age estimate for \systemt based on the mass, radius and model isochrones suggests a significantly younger age than was determined from SED fitting of the host star, we used both kinematics and rotation to obtain alternative age estimates for the host star. 

We followed the method of \citet{discprob} to determine a kinematic age using the our radial velocity and the $Gaia$ proper motions and parallax. We used the equations in \citet{discprob} to calculate the relative thick-to-thin disc, thick-to-halo and thin-to-halo probabilities for \systemAnospace. We used the values in \citet{discprob} for the U$_{asym}$ and V$_{asym}$ asymmetric velocities, dispersion velocities and observed fractions of the thick disc, thin disc and halo populations within the solar neighbourhood and the local standard of rest. We calculated Galactic \textit{U}, \textit{V} and \textit{W} space velocities for \systemA of $-18.03 \pm 0.41$ kms$^{-1}$, $-12.14 \pm 0.67$ kms$^{-1}$ and $12.61 \pm 0.63$ kms$^{-1}$, adjusting for the local standard of rest. We follow the convention of U as positive towards the Galactic centre, V as positive in the direction of Galactic rotation and W as positive towards the North Galactic Pole. We then calculated the relative probabilities for thick-to-thin disc (P$_{\rm thick}$/P$_{\rm thin}$=0.0127), thick-to-halo (P$_{\rm thick}$/P$_{\rm halo}$=7758.6174) and thin-to-halo (P$_{\rm thin}$/P$_{\rm halo}$=609495.1671). 

\citet{discprob} define three interpretations of the thick-to-thin disc probabilities. Firstly, a value of 2 or more means an object is more than likely part of the thick disc, with an average age of ~10 Gyrs \citep{discages}. Secondly, a value of less than 0.5 means an object is more than likely part of the thin disc population, with an average age of $\sim4-5$ Gyrs \citep{discages}. The third population is labelled 'in-between stars' and is when the probability value falls between 0.5 and 2.  With P$_{\rm thick}$/P$_{\rm thin}$=0.0127, which is less than 0.5, this means \systemA is more than likely part of the thin disc population. This gives an age estimate of 4-5~Gyr, which is close to the minimum mass given by the SED fitting.


An additional estimate of the system age was be obtained for the rotational properties of the host star, together with empirical rotation-age (i.e., gyrochronology) relations. Using the stellar radius determined from the SED fitting together with the spectroscopically determined $v\sin i$, we obtain a measure of the star's projected rotation period, $P_{\rm rot} / \sin i = 20.3 \pm 2.3$~d. Using the empirical gyrochronology relations of \citet{mamajek}, we obtain an age estimate of $3.3 \pm 0.7$~Gyr, younger than is determined in Section \ref{sec:host}, but broadly consistent with the kinematic age determination, at the upper end of the values.

All three age estimations for the host star overlap  between 4 and 6~Gyr, with gyrochonology giving the youngest age and the SED fitting the oldest. All three age estimations are not consistent with the age determined using isochrone fitting for the brown dwarf.  This leaves unanswered questions regarding the system. The first scenario is that the brown dwarf did not form at the same time as the host star, although this seems unlikely considering the relatively short orbital period, although the high eccentricity could suggest the brown dwarf was captured. The second scenario is that the brown dwarf isochrones are incorrect.  
The third scenario is that the radius of the brown dwarf is inflated, perhaps an effect due to the intense heating of the atmosphere at periastron, or it could indicate signs of radius scatter in the high-mass brown dwarf regime as is seen for late M dwarfs (e.g. \citealt{parsons18}).

\subsection{Inflation or scatter?}\label{sec:rad}

Using the age estimate from model brown dwarf isochrones from \citet{baraffe03,baraffe15} and \citet{marley21}, our age estimate for the brown dwarf is younger than the ages determined from all other tests we performed on the host star. However, with \systemt lying closer to the 1~Gyr lines than the expected 5 to 10~Gyr lines as determined from stellar and kinematic modelling of the host star, we must consider the possibility of \systemt being inflated. 

\systemt having an inflated radius is unexpected due to its high mass \citep{casewell20}. With \systemtnospace's close approach at periastron of $\sim0.06$~AU, it experiences high levels of irradiation from \systemAnospace. This is reflected in its irradiation temperature, which varies by as much as $\sim1000$~K across the orbit. This is explored more in section \ref{sec:atmo}.

Models have predicted that high metallicity can increase the radius of a brown dwarf \citep{burrows11}, as shown in the isochrones in Figure \ref{fig:massrad}. \systemA has an estimated \feh of $0.32\pm0.05$ dex, meaning it is important to compare to more metal rich models (seen in Figure \ref{fig:massrad} as dotted lines, with the colours representing different ages of the brown dwarf evolution). As can be seen in this figure, increasing \feh to 0.5 dex increases the predicted radius as the mass increases. \systemt lies  on the 1 Gyr \feh $= +0.5$ dex isochrone line. This means that while  higher metallicity models do predict a larger brown dwarf radius,  the radius increase is not large enough to account for the inflated radius we measure.

For low mass M~dwarfs, closer to our object in mass than many brown dwarfs, there is a known radius scatter. \citet{parsons18} find that only 25 per cent of the M dwarfs they observed had radii consistent with models, while 12 per cent were inflated. They also found the host stars in their sample had strong magnetic fields. Whether this scatter continues into the brown dwarf regime is currently debated (e.g. \citealt{19b, casewell20, schaffenroth}). We do not find signs of strong magnetic activity in the lightcurve of \systemAnospace, and therefore the cause for the inflated radius in \systemt is unclear.

\subsection{Irradiation temperature and atmospheric modelling}\label{sec:atmo}

\begin{figure}
    \centering
    \includegraphics[width=\linewidth]{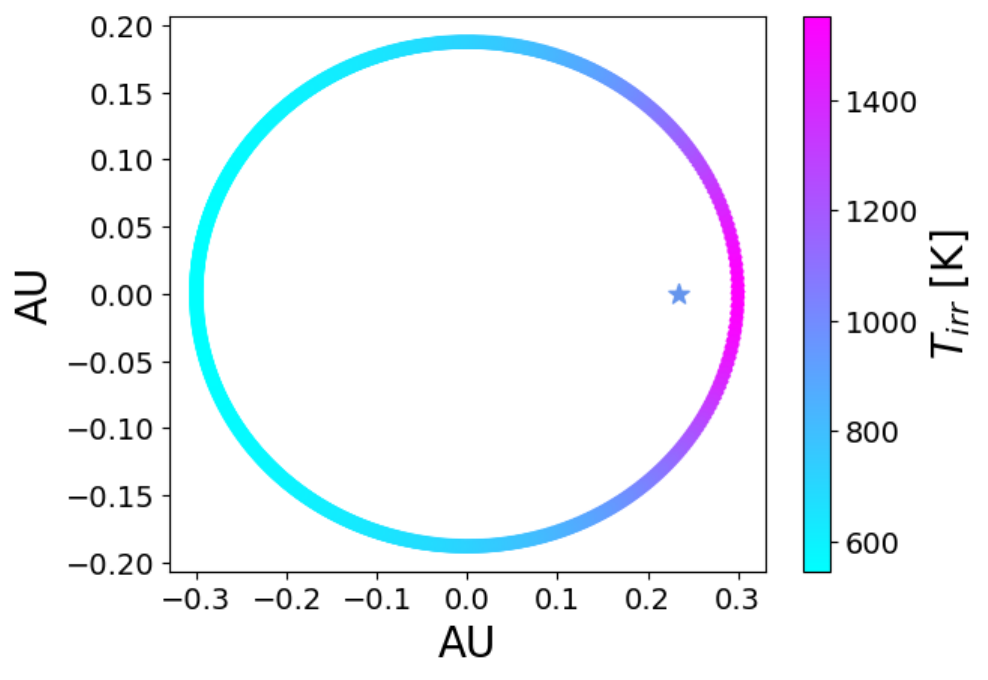}
    \caption{Basic orbit plot for \systemt where the colour represents the irradiation temperature, calculated with Equation \eqref{eq:tirr} from \citet{tsai23}. This ignores the intrinsic internal temperature of the brown dwarf.}
    \label{fig:tirr}
\end{figure}

Due to the extreme nature of \systemtnospace's orbital eccentricity, the temperature change experienced by \systemt as it progresses through its orbit will also change drastically.  We estimate the effective temperature for \systemt as if it were an isolated brown dwarf to be around 1800~K from the 1 Gyr, \feh +0.5 Sonora Bobcat models \citep{marley21}, as these models best fit the mass and radius of the companion. We calculated the irradiation temperature of \systemt as it progresses through its orbit using Equation \eqref{eq:tirr} taken from \citet{tsai23} and create a basic irradiation temperature map of the orbit (Figure \ref{fig:tirr}),

\begin{equation}\label{eq:tirr}
    T_{irr} = T_{\star}\sqrt{\frac{R_{\star}}{D}} ,
\end{equation}
where $T_{irr}$ is the irradiation temperature, $R_{\star}$ is the stellar radius and $D$ is the instantaneous orbital distance.

\begin{figure*}
    \centering
    \includegraphics[width=\linewidth]{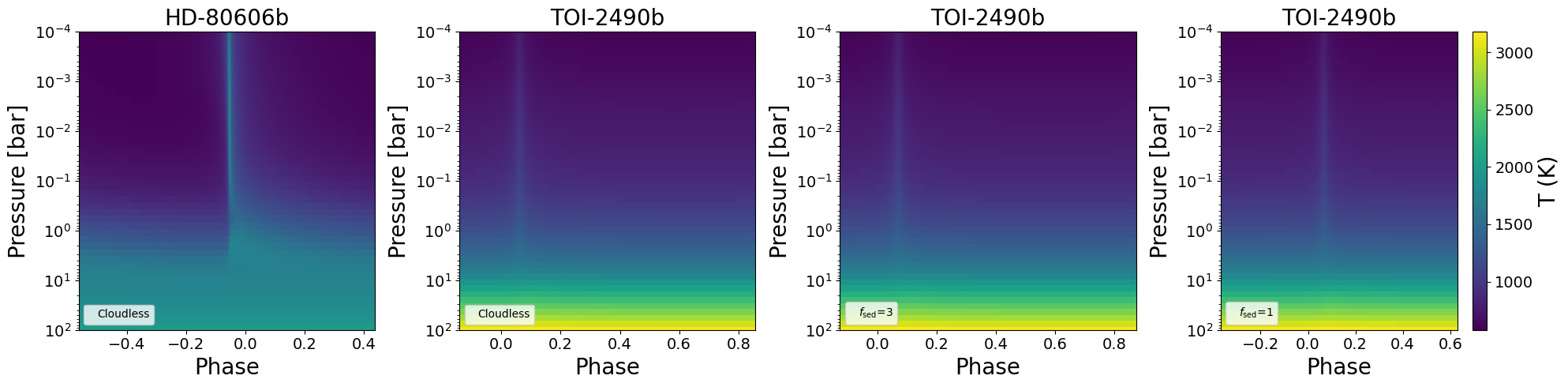}
    \caption{Models generated using \textsc{egp+} showing how the temperature of the atmosphere of \systemt is predicted to change across 3 orbits at different pressure levels. Panels 2, 3 and 4 show the different $f_{sed}$ values tested. The first panel shows the cloud free model of HD~80606b (\citet{mayorga21}) for comparison.}
    \label{fig:pt_hd}
\end{figure*}

Using Equation \eqref{eq:tirr}, we estimate the irradiation temperature for \systemt to be 1552~K at periastron and 545~K at apastron. This means \systemt experiences a 1000~K change in temperature from closest to furthest approach in its orbit. This can be compared to HD~80606b, a $4.16_{-0.005}^{+0.005}$ \mjupnospace, $1.03 \pm 0.02$ \rjup hot Jupiter with a 111.43~d period \citep{pearson22}. It has an orbital eccentricity of 0.93 and semi-major axis of 0.449~AU \citep{tsai23}. HD~80606b is similar to \systemt in that it has a high eccentricity, long period and a large irradiation temperature change, orbiting a star with a similar effective temperature (5645~K, \citealt{tsai23}) to \systemAnospace. \citet{tsai23} calculate that HD~80606b fluctuates in irradiation temperature between 2230~K and 415~K at periastron and apastron respectively. While this change is much larger, the internal temperature of HD~80606b is predicted to be $\sim190$~K, in comparison to that of \systemtnospace's predicted internal temperature of $\sim$1500~K as predicted by \citet{mayorga21} from the \citet{sonora18} models. Comparing the irradiating flux, we determine 16694 Wm$^{-2}$ for HD~80606b and 23855 Wm$^{-2}$ for \systemt.

In order to compare \systemt to HD~80606b, we performed atmospheric modelling using the \citet{mayorga21} one-dimensional, vertical, time-stepping atmospheric code, \textsc{egp+}. \textsc{egp+} is based on the 1D radiative-convection equilibrium models by \citet{marley96,marley02}. \textsc{egp+} tracks the radiative and convective heat input and output throughout the atmosphere, splitting it into pressure layers \citep{mayorga21, robinson14}. \citet{mayorga21} updated the \citet{robinson14} code to include the absorption of incident stellar flux, accounting for a time-dependent distance from the host star \citep{mayorga21}. The updates from \citet{mayorga21} imply \textsc{egp+} is able to simulate the atmospheric response to intense heating at periastron of exoplanets on eccentric orbits. This makes it suitable for analysing the atmosphere of objects such as \systemtnospace.

In Figure \ref{fig:pt_hd}, we show the cloud-free model of HD~80606b, previously published in \citet{mayorga21}, which can be compared to the cloud free model of \systemtnospace. As can be seen in Figure \ref{fig:pt_hd}, during periastron passage \systemt does not experience a dramatic change in temperature at the top of the atmosphere, with only a slightly higher change in temperature, as you progress deeper into the atmosphere. This is the same for both cloudy models. In comparison, HD~80606b experiences a sharp, larger change in temperature throughout the atmosphere than \systemtnospace. \systemt is also clearly much hotter than HD~80606b, as expected.

As is shown in \citet{mayorga21}, a combination of distance from the star at periastron and the companion's gravity will affect the thermal response of the companion atmosphere. \systemtnospace's much higher gravity and slightly further distance at periastron than HD~80606b means its thermal response is much less dramatic, which results in the model predictions in Figure \ref{fig:pt_hd}. In addition, the internal temperature of HD~80606b is predicted to be $\sim190$~K whereas \systemt is predicted to have a much higher internal temperature of $\sim$1500~K. Therefore, while the orbital dynamics of HD~80606b and \systemt may be similar, it is not expected that their atmospheres would react in the same manner to the irradiation.

\begin{figure}
    \centering
    \includegraphics[width=\linewidth]{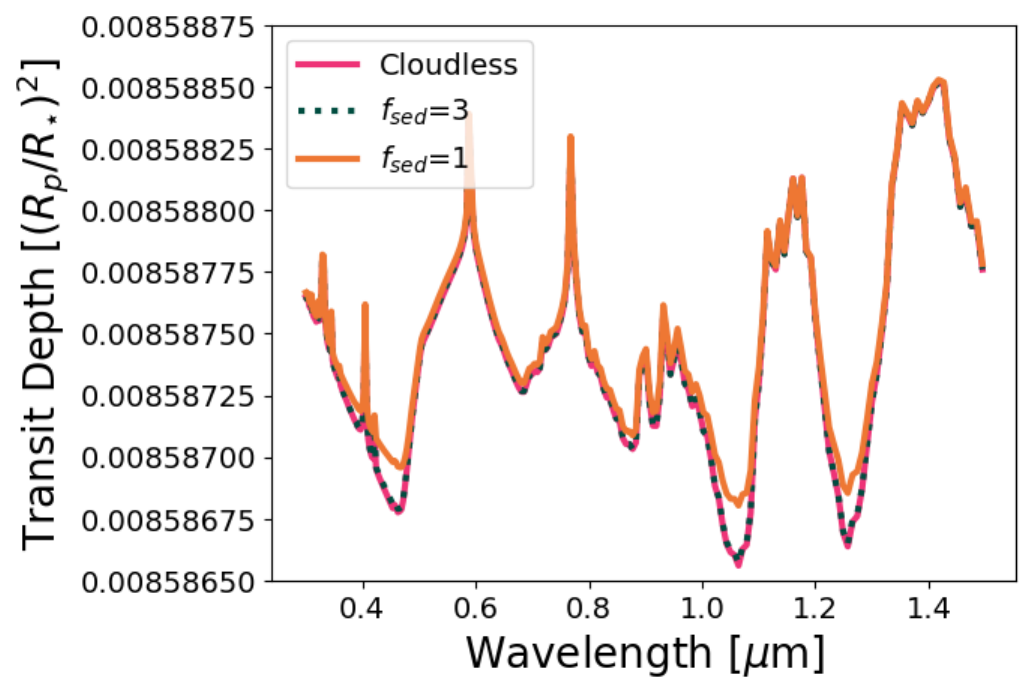}
    \caption{Transmission spectra of \systemt showing the predicted transit depth across a wide wavelength range for cloudless (magenta), $f_{sed}=3$ (teal, dotted) and $f_{sed}=1$ (orange). Changing from cloudless to $f_{sed}=3$ barely changes the transmission depth, but $f_{sed}=1$ has a slightly shallower transit depth.}
    \label{fig:trans_spec}
\end{figure}

For \systemtnospace, we also tested the effect that clouds would have on the transit depth within the \tess and \ngts bandpasses. \textsc{egp+} includes clouds by changing the sedimentation efficiency parameter, \fsednospace. The \fsed value indicates the vertical extent of the cloud, with values above \fsed=1 indicating vertically compressed cloud structures and values less than \fsed=1 indicating vertically extended cloud structures (i.e., the larger the \fsed value, the shallower the cloud depth) \citep{fsed_exp}. Figure \ref{fig:trans_spec} shows the effect clouds have on the transit depth across a wide wavelength range. The cloudless models (magenta) are almost indistinguishable from the \fsednospace=3 cloudy model (teal), with only a slight further difference when plotting the $f_{sed}=1$ cloudy model (orange). 

To compare these spectra with our lightcurves we convolved them with the \ngts \citep{ngts3Ab,NGTS} and \tess \citep{tess} bandpasses to determine the predicted transit depths for each case. For cloudless models, we obtain values of 8.5875 ppt and 8.5874 ppt for \ngts and \tess respectively. For f$_{sed}=3$ models, we get 8.5875 ppt and 8.5874 ppt and for f$_{sed}=1$ models, we get 8.5875 ppt and 8.5874 ppt for \ngts and \tess respectively. There is no difference between the cloudless and cloudy models, to 7 decimal places within each bandpass, and only a 1 $\times 10^{-7}$ difference between the similar filters of \ngts and \tessnospace. We also compared this result to our fitted transit depths from \textsc{allesfitter} of $9.72^{+0.26}_{-0.22}$ ppt, $9.61^{+0.38}_{-0.36}$ ppt and $9.59^{+0.41}_{-0.38}$ ppt for \ngtsnospace, \tess Sector 32 and \tess Sector 5 respectively. The actual \ngts transit depth has a 6$\sigma$ difference from the predicted transit depth, but both \tess sectors have transit depths which are within 3$\sigma$ of the predicted depths. This means our models have predicted the \tess transit depths well but the \ngts transit depths are slightly underestimated.

Additional observations may allow us to understand why the \ngts transits are poorly replicated by the models compared to  the \tess transits, although this effect may simply be due to the atmospheric effects due to \ngts observing from the ground. Observations over a wider wavelength range may also allow us to distinguish between the \fsednospace=3 and cloudless models, as the differences are predicted to be largest in the water bands at $\sim$1.1 and $\sim$1.3 microns, and at $\sim$0.5 microns.

\section{Conclusions and future observations}

In this paper, we report the discovery of the most eccentric transiting brown dwarf found in the desert. \systemt is a $73.6_{-2.3}^{+2.4}$ \mjupnospace, $1.00\pm0.02$ \rjup brown dwarf on a 60.33 d orbit with an eccentricity of $0.77989\pm0.00049$. Using stellar modelling we have found the age to be $7.9_{-1.8}^{+1.5}$ Gyr which is consistent with age estimates from kinematics (4-5~Gyr) and gyrochronology (3.3$\pm$0.7 Gyr)  but is older than estimated using model isochrones for the brown dwarf. This discrepancy could be due to radius inflation of \systemtnospace, or (along with other high-mass inflated objects) could indicate a radius-scatter in the high-mass brown dwarf regime, as seen in low mass M~dwarfs \citep{parsons18}.  

We tested the circularisation timescale for \systemt and checked if the outer object with similar parallax, $\mu$RA and $\mu$Dec (TIC77437539) is able to have caused Kozai-Lidov perturbations on a reasonable timescale. We concluded that it likely formed with this high level of eccentricity. This is consistent with \systemt having formed via stellar formation mechanisms and is likely to have a similar metallicity and age to \systemAnospace. This means the discrepancy between the age of \systemA from stellar and kinematic modelling and the age of \systemt from model isochrones, is likely due to a highly irradiated atmosphere and radius inflation or could indicate a radius scatter in high-mass brown dwarfs, as seen in very low mass M dwarfs. Which of these is the cause of the age discrepancy is unclear.

We studied the irradiation temperature of \systemt as it travels round its orbit and found it changes by $\sim1000$~K. We modelled the thermal response of \systemtnospace's atmosphere and compared it to HD~80606b, due to their similarly drastic orbital dynamics. We found that \systemt has a much less dramatic thermal response than HD~80606b, and concluded it is due to a combination of \systemt having a larger closest approach and higher gravity than HD~80606b. 

\section*{Acknowledgements}

For the purpose of open access, the author has applied a Creative Commons Attribution (CC BY) licence to the Author Accepted Manuscript version arising from this submission.

Based on data collected under the \ngts project at the ESO La Silla Paranal Observatory. The \ngts facility is operated by the consortium institutes with support from the UK Science and Technology Facilities Council (STFC) under projects ST/M001962/1, ST/S002642/1 and ST/W003163/1. 

This study is based on observations collected at the European Southern Observatory.

We acknowledge the use of public TESS data from pipelines at the TESS Science Office and at the TESS Science Processing Operations Center. Funding for the TESS mission is provided by the NASA Explorer Program.  This research has made use of the Exoplanet Follow-up Observation Program website, which is operated by the California Institute of Technology, under contract with the National Aeronautics and Space Administration under the Exoplanet Exploration Program. Resources supporting this work were provided by the NASA High-End Computing (HEC) Program through the NASA Advanced Supercomputing (NAS) Division at Ames Research Center for the production of the SPOC data products.

This work has made use of data from the European Space Agency (ESA) mission \gaia (\url{https://www.cosmos.esa.int/gaia}), processed by the \gaia Data Processing and Analysis Consortium (DPAC, \url{https://www.cosmos.esa.int/web/gaia/dpac/consortium}). Funding for the DPAC has been provided by national institutions, in particular the institutions participating in the \gaia Multilateral Agreement.

This publication makes use of data products from the Two Micron All Sky Survey, which is a joint project of the University of Massachusetts and the Infrared Processing and Analysis Center/California Institute of Technology, funded by the National Aeronautics and Space Administration and the National Science Foundation.

Funding for the SDSS and SDSS-II has been provided by the Alfred P. Sloan Foundation, the Participating Institutions, the National Science Foundation, the U.S. Department of Energy, the National Aeronautics and Space Administration, the Japanese Monbukagakusho, the Max Planck Society, and the Higher Education Funding Council for England. The SDSS Web Site is http://www.sdss.org/. The SDSS is managed by the Astrophysical Research Consortium for the Participating Institutions. The Participating Institutions are the American Museum of Natural History, Astrophysical Institute Potsdam, University of Basel, University of Cambridge, Case Western Reserve University, University of Chicago, Drexel University, Fermilab, the Institute for Advanced Study, the Japan Participation Group, Johns Hopkins University, the Joint Institute for Nuclear Astrophysics, the Kavli Institute for Particle Astrophysics and Cosmology, the Korean Scientist Group, the Chinese Academy of Sciences (LAMOST), Los Alamos National Laboratory, the Max-Planck-Institute for Astronomy (MPIA), the Max-Planck-Institute for Astrophysics (MPA), New Mexico State University, Ohio State University, University of Pittsburgh, University of Portsmouth, Princeton University, the United States Naval Observatory, and the University of Washington.

BH is supported by an STFC studentship.
SLC acknowledges support from an STFC Ernest Rutherford Fellowship (ST/R003726/1).
AJ and RB acknowledge support from ANID -- Millennium  Science
Initiative -- ICN12\_009. 
AJ acknowledges support from FONDECYT project 1210718.
EG gratefully acknowledges support from the UK Science and Technology Facilities Council (STFC; project reference ST/W001047/1). DJA was funded in whole or in part by the UKRI, (Grants ST/X001121/1, EP/X027562/1). PJW and SG acknowledge support from STFC consolidated grants (ST/T000406/1 and ST/X001121/1). 

ML acknowledges support of the Swiss National Science Foundation under grant number PCEFP2\_194576. The contribution of ML has been carried out within the framework of the NCCR PlanetS supported by the Swiss National Science Foundation under grants 51NF40\_182901 and 51NF40\_205606.

JSJ acknowledges support by FONDECYT grant 1201371 and from the ANID BASAL project FB210003.

TT acknowledges support by the DFG Research Unit FOR 2544 "Blue Planets around Red Stars" project No. KU 3625/2-1. TT further acknowledges support by the BNSF program "VIHREN-2021" project No. KP-06-DV/5.

DD acknowledges support from the NASA Exoplanet Research Program grant 18-2XRP18\_2-0136, and from the TESS Guest Investigator Program grant 80NSSC22K1353.

MTP. acknowledges support by FONDECYT-ANID Post-doctoral fellowship no. 3210253.

DR was supported by NASA under award number NNA16BD14C for NASA Academic Mission Services.
\section*{Data Availability}

The \tess data is available via the MAST (MikulskiArchive for Space Telescopes) portal at \url{https://mast.stsci.edu/portal/Mashup/Clients/Mast/Portal.html}.

Public NGTS data is available in the ESO archive.

The other data within this article will be shared on reasonable request to the corresponding author.



\bibliographystyle{mnras}
\bibliography{bibliography.bib} 

\begin{thebibliography}{}
\makeatletter
\relax
\def\mn@urlcharsother{\let\do\@makeother \do\$\do\&\do\#\do\^\do\_\do\%\do\~}
\def\mn@doi{\begingroup\mn@urlcharsother \@ifnextchar [ {\mn@doi@}
  {\mn@doi@[]}}
\def\mn@doi@[#1]#2{\def\@tempa{#1}\ifx\@tempa\@empty \href
  {http://dx.doi.org/#2} {doi:#2}\else \href {http://dx.doi.org/#2} {#1}\fi
  \endgroup}
\def\mn@eprint#1#2{\mn@eprint@#1:#2::\@nil}
\def\mn@eprint@arXiv#1{\href {http://arxiv.org/abs/#1} {{\tt arXiv:#1}}}
\def\mn@eprint@dblp#1{\href {http://dblp.uni-trier.de/rec/bibtex/#1.xml}
  {dblp:#1}}
\def\mn@eprint@#1:#2:#3:#4\@nil{\def\@tempa {#1}\def\@tempb {#2}\def\@tempc
  {#3}\ifx \@tempc \@empty \let \@tempc \@tempb \let \@tempb \@tempa \fi \ifx
  \@tempb \@empty \def\@tempb {arXiv}\fi \@ifundefined
  {mn@eprint@\@tempb}{\@tempb:\@tempc}{\expandafter \expandafter \csname
  mn@eprint@\@tempb\endcsname \expandafter{\@tempc}}}

\bibitem[\protect\citeauthoryear{Acton et~al.,}{Acton et~al.}{2021}]{19b}
Acton J.~S.,  et~al., 2021, \mn@doi [\mnras] {10.1093/mnras/stab1459}, 505,
  2741–2752

\bibitem[\protect\citeauthoryear{Armitage}{Armitage}{2020}]{kozailidov}
Armitage P.~J.,  2020, Early Evolution of Planetary Systems, 2 edn.
Cambridge University Press, p. 247–300, \mn@doi{10.1017/9781108344227.008}

\bibitem[\protect\citeauthoryear{{Artigau} et~al.,}{{Artigau}
  et~al.}{2021}]{toi1278b}
{Artigau} {\'E}.,  et~al., 2021, \mn@doi [\aj] {10.3847/1538-3881/ac096d},
  \href {https://ui.adsabs.harvard.edu/abs/2021AJ....162..144A} {162, 144}

\bibitem[\protect\citeauthoryear{{Babusiaux} et~al.,}{{Babusiaux}
  et~al.}{2023}]{gaiaextra}
{Babusiaux} C.,  et~al., 2023, \mn@doi [\aap] {10.1051/0004-6361/202243790},
  \href {https://ui.adsabs.harvard.edu/abs/2023A&A...674A..32B} {674, A32}

\bibitem[\protect\citeauthoryear{{Bakos} et~al.,}{{Bakos}
  et~al.}{2009}]{hatp13c}
{Bakos} G.~{\'A}.,  et~al., 2009, \mn@doi [\apj] {10.1088/0004-637X/707/1/446},
  \href {https://ui.adsabs.harvard.edu/abs/2009ApJ...707..446B} {707, 446}

\bibitem[\protect\citeauthoryear{{Baraffe}, {Chabrier}, {Allard}  \&
  {Hauschildt}}{{Baraffe} et~al.}{2002}]{baraffe02}
{Baraffe} I.,  {Chabrier} G.,  {Allard} F.,   {Hauschildt} P.~H.,  2002,
  \mn@doi [\aap] {10.1051/0004-6361:20011638}, \href
  {https://ui.adsabs.harvard.edu/abs/2002A&A...382..563B} {382, 563}

\bibitem[\protect\citeauthoryear{{Baraffe}, {Chabrier}, {Barman}, {Allard}  \&
  {Hauschildt}}{{Baraffe} et~al.}{2003}]{baraffe03}
{Baraffe} I.,  {Chabrier} G.,  {Barman} T.~S.,  {Allard} F.,   {Hauschildt}
  P.~H.,  2003, \mn@doi [\aap] {10.1051/0004-6361:20030252}, \href
  {https://ui.adsabs.harvard.edu/abs/2003A&A...402..701B} {402, 701}

\bibitem[\protect\citeauthoryear{{Baraffe}, {Homeier}, {Allard}  \&
  {Chabrier}}{{Baraffe} et~al.}{2015}]{baraffe15}
{Baraffe} I.,  {Homeier} D.,  {Allard} F.,   {Chabrier} G.,  2015, \mn@doi
  [\aap] {10.1051/0004-6361/201425481}, \href
  {https://ui.adsabs.harvard.edu/abs/2015A&A...577A..42B} {577, A42}

\bibitem[\protect\citeauthoryear{{Bate}, {Bonnell}  \& {Bromm}}{{Bate}
  et~al.}{2002}]{bate02}
{Bate} M.~R.,  {Bonnell} I.~A.,   {Bromm} V.,  2002, \mn@doi [MNRAS]
  {10.1046/j.1365-8711.2002.05539.x}, \href
  {https://ui.adsabs.harvard.edu/abs/2002MNRAS.332L..65B} {332, L65}

\bibitem[\protect\citeauthoryear{{Bayliss} et~al.,}{{Bayliss}
  et~al.}{2017}]{epic201702477b}
{Bayliss} D.,  et~al., 2017, \mn@doi [\aj] {10.3847/1538-3881/153/1/15}, \href
  {https://ui.adsabs.harvard.edu/abs/2017AJ....153...15B} {153, 15}

\bibitem[\protect\citeauthoryear{{Beatty} et~al.,}{{Beatty}
  et~al.}{2007}]{hat205}
{Beatty} T.~G.,  et~al., 2007, \mn@doi [\apj] {10.1086/518413}, \href
  {https://ui.adsabs.harvard.edu/abs/2007ApJ...663..573B} {663, 573}

\bibitem[\protect\citeauthoryear{{Beatty} et~al.,}{{Beatty}
  et~al.}{2014}]{beatty14_1b}
{Beatty} T.~G.,  et~al., 2014, \mn@doi [\apj] {10.1088/0004-637X/783/2/112},
  \href {https://ui.adsabs.harvard.edu/abs/2014ApJ...783..112B} {783, 112}

\bibitem[\protect\citeauthoryear{{Beatty}, {Morley}, {Curtis}, {Burrows},
  {Davenport}  \& {Montet}}{{Beatty} et~al.}{2018}]{beatty18}
{Beatty} T.~G.,  {Morley} C.~V.,  {Curtis} J.~L.,  {Burrows} A.,  {Davenport}
  J. R.~A.,   {Montet} B.~T.,  2018, \mn@doi [\aj] {10.3847/1538-3881/aad697},
  \href {https://ui.adsabs.harvard.edu/abs/2018AJ....156..168B} {156, 168}

\bibitem[\protect\citeauthoryear{{Benni} et~al.,}{{Benni} et~al.}{2021}]{gpx1b}
{Benni} P.,  et~al., 2021, \mn@doi [\mnras] {10.1093/mnras/stab1567}, \href
  {https://ui.adsabs.harvard.edu/abs/2021MNRAS.505.4956B} {505, 4956}

\bibitem[\protect\citeauthoryear{{Bensby}, {Feltzing}  \& {Oey}}{{Bensby}
  et~al.}{2014}]{discprob}
{Bensby} T.,  {Feltzing} S.,   {Oey} M.~S.,  2014, \mn@doi [\aap]
  {10.1051/0004-6361/201322631}, \href
  {https://ui.adsabs.harvard.edu/abs/2014A&A...562A..71B} {562, A71}

\bibitem[\protect\citeauthoryear{Best, Dupuy, Liu, Siverd  \& Zhang}{Best
  et~al.}{2021}]{ultracool}
Best W. M.~J.,  Dupuy T.~J.,  Liu M.~C.,  Siverd R.~J.,   Zhang Z.,  2021, {The
  UltracoolSheet: Photometry, Astrometry, Spectroscopy, and Multiplicity for
  3000+ Ultracool Dwarfs and Imaged Exoplanets},
  \mn@doi{10.5281/zenodo.4570814}, \url
  {https://doi.org/10.5281/zenodo.4570814}

\bibitem[\protect\citeauthoryear{{Bonomo} et~al.,}{{Bonomo}
  et~al.}{2015}]{kepler39b}
{Bonomo} A.~S.,  et~al., 2015, \mn@doi [\aap] {10.1051/0004-6361/201323042},
  \href {https://ui.adsabs.harvard.edu/abs/2015A&A...575A..85B} {575, A85}

\bibitem[\protect\citeauthoryear{{Bouchy} et~al.,}{{Bouchy}
  et~al.}{2011}]{corot15b}
{Bouchy} F.,  et~al., 2011, \mn@doi [\aap] {10.1051/0004-6361/201015276}, \href
  {https://ui.adsabs.harvard.edu/abs/2011A&A...525A..68B} {525, A68}

\bibitem[\protect\citeauthoryear{{Bowler}, {Blunt}  \& {Nielsen}}{{Bowler}
  et~al.}{2020}]{bowler20}
{Bowler} B.~P.,  {Blunt} S.~C.,   {Nielsen} E.~L.,  2020, \mn@doi [\aj]
  {10.3847/1538-3881/ab5b11}, \href
  {https://ui.adsabs.harvard.edu/abs/2020AJ....159...63B} {159, 63}

\bibitem[\protect\citeauthoryear{{Brahm}, {Jord{\'a}n}  \& {Espinoza}}{{Brahm}
  et~al.}{2017a}]{ceres}
{Brahm} R.,  {Jord{\'a}n} A.,   {Espinoza} N.,  2017a, \mn@doi [\pasp]
  {10.1088/1538-3873/aa5455}, \href
  {https://ui.adsabs.harvard.edu/abs/2017PASP..129c4002B} {129, 034002}

\bibitem[\protect\citeauthoryear{{Brahm}, {Jord{\'a}n}, {Hartman}  \&
  {Bakos}}{{Brahm} et~al.}{2017b}]{zaspe}
{Brahm} R.,  {Jord{\'a}n} A.,  {Hartman} J.,   {Bakos} G.,  2017b, \mn@doi
  [\mnras] {10.1093/mnras/stx144}, \href
  {https://ui.adsabs.harvard.edu/abs/2017MNRAS.467..971B} {467, 971}

\bibitem[\protect\citeauthoryear{{Brahm} et~al.,}{{Brahm}
  et~al.}{2019}]{hd1397}
{Brahm} R.,  et~al., 2019, \mn@doi [\aj] {10.3847/1538-3881/ab279a}, \href
  {https://ui.adsabs.harvard.edu/abs/2019AJ....158...45B} {158, 45}

\bibitem[\protect\citeauthoryear{Burgasser}{Burgasser}{2008}]{burgasser}
Burgasser A.~J.,  2008, Physics Today, 61, 70

\bibitem[\protect\citeauthoryear{{Burrows} \& {Liebert}}{{Burrows} \&
  {Liebert}}{1993}]{burrows93}
{Burrows} A.,  {Liebert} J.,  1993, \mn@doi [Reviews of Modern Physics]
  {10.1103/RevModPhys.65.301}, \href
  {https://ui.adsabs.harvard.edu/abs/1993RvMP...65..301B} {65, 301}

\bibitem[\protect\citeauthoryear{{Burrows}, {Heng}  \& {Nampaisarn}}{{Burrows}
  et~al.}{2011}]{burrows11}
{Burrows} A.,  {Heng} K.,   {Nampaisarn} T.,  2011, \mn@doi [\apj]
  {10.1088/0004-637X/736/1/47}, \href
  {https://ui.adsabs.harvard.edu/abs/2011ApJ...736...47B} {736, 47}

\bibitem[\protect\citeauthoryear{{Caldwell} et~al.,}{{Caldwell}
  et~al.}{2020}]{caldwell20}
{Caldwell} D.~A.,  et~al., 2020, \mn@doi [Research Notes of the American
  Astronomical Society] {10.3847/2515-5172/abc9b3}, \href
  {https://ui.adsabs.harvard.edu/abs/2020RNAAS...4..201C} {4, 201}

\bibitem[\protect\citeauthoryear{Carmichael}{Carmichael}{2022}]{imporvedradii}
Carmichael T.~W.,  2022, \mn@doi [Monthly Notices of the Royal Astronomical
  Society] {10.1093/mnras/stac3720}, 519, 5177

\bibitem[\protect\citeauthoryear{{Carmichael}, {Latham}  \&
  {Vanderburg}}{{Carmichael} et~al.}{2019}]{carmichael19}
{Carmichael} T.~W.,  {Latham} D.~W.,   {Vanderburg} A.~M.,  2019, \mn@doi [\aj]
  {10.3847/1538-3881/ab245e}, \href
  {https://ui.adsabs.harvard.edu/abs/2019AJ....158...38C} {158, 38}

\bibitem[\protect\citeauthoryear{{Carmichael} et~al.,}{{Carmichael}
  et~al.}{2020}]{carmichael20}
{Carmichael} T.~W.,  et~al., 2020, \mn@doi [\aj] {10.3847/1538-3881/ab9b84},
  \href {https://ui.adsabs.harvard.edu/abs/2020AJ....160...53C} {160, 53}

\bibitem[\protect\citeauthoryear{Carmichael et~al.,}{Carmichael
  et~al.}{2021}]{carmichael21}
Carmichael T.~W.,  et~al., 2021, \mn@doi [\aj] {10.3847/1538-3881/abd4e1}, 161,
  97

\bibitem[\protect\citeauthoryear{{Carmichael} et~al.,}{{Carmichael}
  et~al.}{2022}]{carmichael22}
{Carmichael} T.~W.,  et~al., 2022, \mn@doi [\mnras] {10.1093/mnras/stac1666},
  \href {https://ui.adsabs.harvard.edu/abs/2022MNRAS.514.4944C} {514, 4944}

\bibitem[\protect\citeauthoryear{{Casewell}, {Debes}, {Braker}, {Cushing},
  {Mace}, {Marley}  \& {Kirkpatrick}}{{Casewell} et~al.}{2020}]{casewell20}
{Casewell} S.~L.,  {Debes} J.,  {Braker} I.~P.,  {Cushing} M.~C.,  {Mace} G.,
  {Marley} M.~S.,   {Kirkpatrick} J.~D.,  2020, \mn@doi [\mnras]
  {10.1093/mnras/staa3184}, \href
  {https://ui.adsabs.harvard.edu/abs/2020MNRAS.499.5318C} {499, 5318}

\bibitem[\protect\citeauthoryear{{Chaturvedi}, {Chakraborty}, {Anandarao},
  {Roy}  \& {Mahadevan}}{{Chaturvedi} et~al.}{2016}]{j2343}
{Chaturvedi} P.,  {Chakraborty} A.,  {Anandarao} B.~G.,  {Roy} A.,
  {Mahadevan} S.,  2016, \mn@doi [\mnras] {10.1093/mnras/stw1560}, \href
  {https://ui.adsabs.harvard.edu/abs/2016MNRAS.462..554C} {462, 554}

\bibitem[\protect\citeauthoryear{Csizmadia}{Csizmadia}{2016}]{NLTTrad}
Csizmadia S.,  2016, in , The {CoRoT} Legacy Book.
{EDP} Sciences, p.~143, \mn@doi{10.1051/978-2-7598-1876-1.c036}, \url
  {https://doi.org/10.1051\%2F978-2-7598-1876-1.c036}

\bibitem[\protect\citeauthoryear{{Csizmadia} et~al.,}{{Csizmadia}
  et~al.}{2015}]{corot33b}
{Csizmadia} S.,  et~al., 2015, \mn@doi [\aap] {10.1051/0004-6361/201526763},
  \href {https://ui.adsabs.harvard.edu/abs/2015A&A...584A..13C} {584, A13}

\bibitem[\protect\citeauthoryear{{David}, {Hillenbrand}, {Gillen}, {Cody},
  {Howell}, {Isaacson}  \& {Livingston}}{{David} et~al.}{2019}]{rik72b}
{David} T.~J.,  {Hillenbrand} L.~A.,  {Gillen} E.,  {Cody} A.~M.,  {Howell}
  S.~B.,  {Isaacson} H.~T.,   {Livingston} J.~H.,  2019, \mn@doi [\apj]
  {10.3847/1538-4357/aafe09}, \href
  {https://ui.adsabs.harvard.edu/abs/2019ApJ...872..161D} {872, 161}

\bibitem[\protect\citeauthoryear{{Deleuil} et~al.,}{{Deleuil}
  et~al.}{2008}]{corot3b}
{Deleuil} M.,  et~al., 2008, \mn@doi [\aap] {10.1051/0004-6361:200810625},
  \href {https://ui.adsabs.harvard.edu/abs/2008A&A...491..889D} {491, 889}

\bibitem[\protect\citeauthoryear{{D{\'\i}az} et~al.,}{{D{\'\i}az}
  et~al.}{2013}]{koi205b}
{D{\'\i}az} R.~F.,  et~al., 2013, \mn@doi [\aap] {10.1051/0004-6361/201321124},
  \href {https://ui.adsabs.harvard.edu/abs/2013A&A...551L...9D} {551, L9}

\bibitem[\protect\citeauthoryear{{D{\'\i}az} et~al.,}{{D{\'\i}az}
  et~al.}{2014}]{koi189b}
{D{\'\i}az} R.~F.,  et~al., 2014, \mn@doi [\aap] {10.1051/0004-6361/201424406},
  \href {https://ui.adsabs.harvard.edu/abs/2014A&A...572A.109D} {572, A109}

\bibitem[\protect\citeauthoryear{{Endl}, {Hatzes}, {Cochran}, {McArthur},
  {Allende Prieto}, {Paulson}, {Guenther}  \& {Bedalov}}{{Endl}
  et~al.}{2004}]{HD137510}
{Endl} M.,  {Hatzes} A.~P.,  {Cochran} W.~D.,  {McArthur} B.,  {Allende Prieto}
  C.,  {Paulson} D.~B.,  {Guenther} E.,   {Bedalov} A.,  2004, \mn@doi [\apj]
  {10.1086/422310}, \href
  {https://ui.adsabs.harvard.edu/abs/2004ApJ...611.1121E} {611, 1121}

\bibitem[\protect\citeauthoryear{{Foreman-Mackey}, {Hogg}, {Lang}  \&
  {Goodman}}{{Foreman-Mackey} et~al.}{2013}]{emcee}
{Foreman-Mackey} D.,  {Hogg} D.~W.,  {Lang} D.,   {Goodman} J.,  2013, \mn@doi
  [\pasp] {10.1086/670067}, \href
  {https://ui.adsabs.harvard.edu/abs/2013PASP..125..306F} {125, 306}

\bibitem[\protect\citeauthoryear{{Fukugita}, {Ichikawa}, {Gunn}, {Doi},
  {Shimasaku}  \& {Schneider}}{{Fukugita} et~al.}{1996}]{sdssphot}
{Fukugita} M.,  {Ichikawa} T.,  {Gunn} J.~E.,  {Doi} M.,  {Shimasaku} K.,
  {Schneider} D.~P.,  1996, \mn@doi [\aj] {10.1086/117915}, \href
  {https://ui.adsabs.harvard.edu/abs/1996AJ....111.1748F} {111, 1748}

\bibitem[\protect\citeauthoryear{{Gaia Collaboration} et~al.,}{{Gaia
  Collaboration} et~al.}{2016}]{gaia}
{Gaia Collaboration} et~al., 2016, \mn@doi [\aap]
  {10.1051/0004-6361/201629272}, \href
  {https://ui.adsabs.harvard.edu/abs/2016A&A...595A...1G} {595, A1}

\bibitem[\protect\citeauthoryear{{Gaia Collaboration} et~al.,}{{Gaia
  Collaboration} et~al.}{2023}]{gaiadr3}
{Gaia Collaboration} et~al., 2023, \mn@doi [\aap]
  {10.1051/0004-6361/202243940}, \href
  {https://ui.adsabs.harvard.edu/abs/2023A&A...674A...1G} {674, A1}

\bibitem[\protect\citeauthoryear{{Gao}, {Marley}  \& {Ackerman}}{{Gao}
  et~al.}{2018}]{fsed_exp}
{Gao} P.,  {Marley} M.~S.,   {Ackerman} A.~S.,  2018, \mn@doi [\apj]
  {10.3847/1538-4357/aab0a1}, \href
  {https://ui.adsabs.harvard.edu/abs/2018ApJ...855...86G} {855, 86}

\bibitem[\protect\citeauthoryear{{Gill} et~al.,}{{Gill} et~al.}{2020a}]{mono}
{Gill} S.,  et~al., 2020a, \mn@doi [\mnras] {10.1093/mnras/stz3212}, \href
  {https://ui.adsabs.harvard.edu/abs/2020MNRAS.491.1548G} {491, 1548}

\bibitem[\protect\citeauthoryear{{Gill} et~al.,}{{Gill}
  et~al.}{2020b}]{tic2310}
{Gill} S.,  et~al., 2020b, \mn@doi [\mnras] {10.1093/mnras/staa1248}, \href
  {https://ui.adsabs.harvard.edu/abs/2020MNRAS.495.2713G} {495, 2713}

\bibitem[\protect\citeauthoryear{{Gill} et~al.,}{{Gill}
  et~al.}{2022}]{tic320687387b}
{Gill} S.,  et~al., 2022, \mn@doi [\mnras] {10.1093/mnras/stac798}, \href
  {https://ui.adsabs.harvard.edu/abs/2022MNRAS.513.1785G} {513, 1785}

\bibitem[\protect\citeauthoryear{{Gillen}, {Hillenbrand}, {David}, {Aigrain},
  {Rebull}, {Stauffer}, {Cody}  \& {Queloz}}{{Gillen} et~al.}{2017}]{ad3116b}
{Gillen} E.,  {Hillenbrand} L.~A.,  {David} T.~J.,  {Aigrain} S.,  {Rebull} L.,
   {Stauffer} J.,  {Cody} A.~M.,   {Queloz} D.,  2017, \mn@doi [\apj]
  {10.3847/1538-4357/aa84b3}, \href
  {https://ui.adsabs.harvard.edu/abs/2017ApJ...849...11G} {849, 11}

\bibitem[\protect\citeauthoryear{{Grether} \& {Lineweaver}}{{Grether} \&
  {Lineweaver}}{2006}]{grether06}
{Grether} D.,  {Lineweaver} C.~H.,  2006, \mn@doi [\apj] {10.1086/500161},
  \href {https://ui.adsabs.harvard.edu/abs/2006ApJ...640.1051G} {640, 1051}

\bibitem[\protect\citeauthoryear{Grieves et~al.,}{Grieves
  et~al.}{2021}]{grieves21}
Grieves N.,  et~al., 2021, \mn@doi [\aap] {10.1051/0004-6361/202141145}, 652,
  A127

\bibitem[\protect\citeauthoryear{{Gunn} et~al.,}{{Gunn} et~al.}{1998}]{sdsscam}
{Gunn} J.~E.,  et~al., 1998, \mn@doi [\aj] {10.1086/300645}, \href
  {https://ui.adsabs.harvard.edu/abs/1998AJ....116.3040G} {116, 3040}

\bibitem[\protect\citeauthoryear{{G{\"u}nther} \& {Daylan}}{{G{\"u}nther} \&
  {Daylan}}{2019}]{allesfitter-code}
{G{\"u}nther} M.~N.,  {Daylan} T.,  2019, {Allesfitter: Flexible Star and
  Exoplanet Inference From Photometry and Radial Velocity}, Astrophysics Source
  Code Library (\mn@eprint {ascl} {1903.003})

\bibitem[\protect\citeauthoryear{{G{\"u}nther} \& {Daylan}}{{G{\"u}nther} \&
  {Daylan}}{2021}]{allesfitter-paper}
{G{\"u}nther} M.~N.,  {Daylan} T.,  2021, \mn@doi [\apjs]
  {10.3847/1538-4365/abe70e}, \href
  {https://ui.adsabs.harvard.edu/abs/2021ApJS..254...13G} {254, 13}

\bibitem[\protect\citeauthoryear{{G{\"u}nther} et~al.,}{{G{\"u}nther}
  et~al.}{2018}]{ngts3Ab}
{G{\"u}nther} M.~N.,  et~al., 2018, \mn@doi [\mnras] {10.1093/mnras/sty1193},
  \href {https://ui.adsabs.harvard.edu/abs/2018MNRAS.478.4720G} {478, 4720}

\bibitem[\protect\citeauthoryear{{Heller}, {Jackson}, {Barnes}, {Greenberg}  \&
  {Homeier}}{{Heller} et~al.}{2010}]{heller10}
{Heller} R.,  {Jackson} B.,  {Barnes} R.,  {Greenberg} R.,   {Homeier} D.,
  2010, \mn@doi [\aap] {10.1051/0004-6361/200912826}, \href
  {https://ui.adsabs.harvard.edu/abs/2010A&A...514A..22H} {514, A22}

\bibitem[\protect\citeauthoryear{{Henderson} et~al.,}{{Henderson}
  et~al.}{2024}]{henderson24}
{Henderson} B.~A.,  et~al., 2024, \mn@doi [\mnras] {10.1093/mnras/stae508},
  \href {https://ui.adsabs.harvard.edu/abs/2024MNRAS.tmp..603H} {}

\bibitem[\protect\citeauthoryear{Hodžić et~al.,}{Hodžić
  et~al.}{2018}]{wasp128b}
Hodžić V.,  et~al., 2018, \mn@doi [\mnras] {10.1093/mnras/sty2512}, 481, 5091

\bibitem[\protect\citeauthoryear{{Irwin} et~al.,}{{Irwin} et~al.}{2010}]{nltt}
{Irwin} J.,  et~al., 2010, \mn@doi [\apj] {10.1088/0004-637X/718/2/1353}, \href
  {https://ui.adsabs.harvard.edu/abs/2010ApJ...718.1353I} {718, 1353}

\bibitem[\protect\citeauthoryear{{Irwin} et~al.,}{{Irwin} et~al.}{2018}]{lp261}
{Irwin} J.~M.,  et~al., 2018, \mn@doi [\aj] {10.3847/1538-3881/aad9a3}, \href
  {https://ui.adsabs.harvard.edu/abs/2018AJ....156..140I} {156, 140}

\bibitem[\protect\citeauthoryear{{Jackman} et~al.,}{{Jackman}
  et~al.}{2019}]{ngts7ab}
{Jackman} J. A.~G.,  et~al., 2019, \mn@doi [\mnras] {10.1093/mnras/stz2496},
  \href {https://ui.adsabs.harvard.edu/abs/2019MNRAS.489.5146J} {489, 5146}

\bibitem[\protect\citeauthoryear{{Jackson}, {Greenberg}  \& {Barnes}}{{Jackson}
  et~al.}{2008}]{jackson08}
{Jackson} B.,  {Greenberg} R.,   {Barnes} R.,  2008, \mn@doi [\apj]
  {10.1086/529187}, \href
  {https://ui.adsabs.harvard.edu/abs/2008ApJ...678.1396J} {678, 1396}

\bibitem[\protect\citeauthoryear{Jeffreys}{Jeffreys}{1998}]{jeffreys98theory}
Jeffreys H.,  1998, The Theory of Probability.
Oxford Classic Texts in the Physical Sciences, OUP Oxford, \url
  {https://books.google.co.uk/books?id=vh9Act9rtzQC}

\bibitem[\protect\citeauthoryear{{Jenkins}}{{Jenkins}}{2002}]{jenkins02}
{Jenkins} J.~M.,  2002, \mn@doi [\apj] {10.1086/341136}, \href
  {http://adsabs.harvard.edu/abs/2002ApJ...575..493J} {575, 493}

\bibitem[\protect\citeauthoryear{{Jenkins} et~al.,}{{Jenkins}
  et~al.}{2009}]{HD191760b}
{Jenkins} J.~S.,  et~al., 2009, \mn@doi [\mnras]
  {10.1111/j.1365-2966.2009.15097.x}, \href
  {https://ui.adsabs.harvard.edu/abs/2009MNRAS.398..911J} {398, 911}

\bibitem[\protect\citeauthoryear{{Jenkins} et~al.,}{{Jenkins}
  et~al.}{2016}]{spoc}
{Jenkins} J.~M.,  et~al., 2016, in {Chiozzi} G.,  {Guzman} J.~C.,  eds,
  Society of Photo-Optical Instrumentation Engineers (SPIE) Conference Series
  Vol. 9913, Software and Cyberinfrastructure for Astronomy IV. p. 99133E,
  \mn@doi{10.1117/12.2233418}

\bibitem[\protect\citeauthoryear{{Jenkins}, {Tenenbaum}, {Seader}, {Burke},
  {McCauliff}, {Smith}, {Twicken}  \& {Chandrasekaran}}{{Jenkins}
  et~al.}{2020}]{jenkins20}
{Jenkins} J.~M.,  {Tenenbaum} P.,  {Seader} S.,  {Burke} C.~J.,  {McCauliff}
  S.~D.,  {Smith} J.~C.,  {Twicken} J.~D.,   {Chandrasekaran} H.,  2020,
  {Kepler Data Processing Handbook: Transiting Planet Search}, Kepler Science
  Document KSCI-19081-003

\bibitem[\protect\citeauthoryear{{Johnson} et~al.,}{{Johnson}
  et~al.}{2011}]{lhs6343}
{Johnson} J.~A.,  et~al., 2011, \mn@doi [\apj] {10.1088/0004-637X/730/2/79},
  \href {https://ui.adsabs.harvard.edu/abs/2011ApJ...730...79J} {730, 79}

\bibitem[\protect\citeauthoryear{Kass \& Raftery}{Kass \& Raftery}{1995}]{kass}
Kass R.~E.,  Raftery A.~E.,  1995, \mn@doi [Journal of the American Statistical
  Association] {10.1080/01621459.1995.10476572}, 90, 773

\bibitem[\protect\citeauthoryear{{Kaufer}, {Stahl}, {Tubbesing},
  {N{\o}rregaard}, {Avila}, {Francois}, {Pasquini}  \& {Pizzella}}{{Kaufer}
  et~al.}{1999}]{feros}
{Kaufer} A.,  {Stahl} O.,  {Tubbesing} S.,  {N{\o}rregaard} P.,  {Avila} G.,
  {Francois} P.,  {Pasquini} L.,   {Pizzella} A.,  1999, The Messenger, \href
  {https://ui.adsabs.harvard.edu/abs/1999Msngr..95....8K} {95, 8}

\bibitem[\protect\citeauthoryear{{Khandelwal} et~al.,}{{Khandelwal}
  et~al.}{2023}]{toi4603b}
{Khandelwal} A.,  et~al., 2023, \mn@doi [\aap] {10.1051/0004-6361/202245608},
  \href {https://ui.adsabs.harvard.edu/abs/2023A&A...672L...7K} {672, L7}

\bibitem[\protect\citeauthoryear{Kipping}{Kipping}{2013}]{kipping13}
Kipping D.~M.,  2013, \mn@doi [\mnras] {10.1093/mnras/stt1435}, 435, 2152

\bibitem[\protect\citeauthoryear{{Latham}, {Mazeh}, {Stefanik}, {Mayor}  \&
  {Burki}}{{Latham} et~al.}{1989}]{rvbd}
{Latham} D.~W.,  {Mazeh} T.,  {Stefanik} R.~P.,  {Mayor} M.,   {Burki} G.,
  1989, \mn@doi [\nat] {10.1038/339038a0}, \href
  {https://ui.adsabs.harvard.edu/abs/1989Natur.339...38L} {339, 38}

\bibitem[\protect\citeauthoryear{Lin et~al.,}{Lin et~al.}{2023}]{lin23}
Lin Z.,  et~al., 2023, \mn@doi [Monthly Notices of the Royal Astronomical
  Society] {10.1093/mnras/stad1745}, 523, 6162

\bibitem[\protect\citeauthoryear{{Lucy} \& {Sweeney}}{{Lucy} \&
  {Sweeney}}{1971}]{Lucy71}
{Lucy} L.~B.,  {Sweeney} M.~A.,  1971, \mn@doi [\aj] {10.1086/111159}, \href
  {https://ui.adsabs.harvard.edu/abs/1971AJ.....76..544L} {76, 544}

\bibitem[\protect\citeauthoryear{Ma \& Ge}{Ma \& Ge}{2014}]{mage14}
Ma B.,  Ge J.,  2014, \mn@doi [\mnras] {10.1093/mnras/stu134}, 439, 2781

\bibitem[\protect\citeauthoryear{{Maldonado} et~al.,}{{Maldonado}
  et~al.}{2023}]{toi5375b}
{Maldonado} J.,  et~al., 2023, \mn@doi [\aap] {10.1051/0004-6361/202346096},
  \href {https://ui.adsabs.harvard.edu/abs/2023A&A...674A.132M} {674, A132}

\bibitem[\protect\citeauthoryear{{Marley}, {Saumon}, {Guillot}, {Freedman},
  {Hubbard}, {Burrows}  \& {Lunine}}{{Marley} et~al.}{1996}]{marley96}
{Marley} M.~S.,  {Saumon} D.,  {Guillot} T.,  {Freedman} R.~S.,  {Hubbard}
  W.~B.,  {Burrows} A.,   {Lunine} J.~I.,  1996, \mn@doi [Science]
  {10.1126/science.272.5270.1919}, \href
  {https://ui.adsabs.harvard.edu/abs/1996Sci...272.1919M} {272, 1919}

\bibitem[\protect\citeauthoryear{{Marley}, {Seager}, {Saumon}, {Lodders},
  {Ackerman}, {Freedman}  \& {Fan}}{{Marley} et~al.}{2002}]{marley02}
{Marley} M.~S.,  {Seager} S.,  {Saumon} D.,  {Lodders} K.,  {Ackerman} A.~S.,
  {Freedman} R.~S.,   {Fan} X.,  2002, \mn@doi [\apj] {10.1086/338800}, \href
  {https://ui.adsabs.harvard.edu/abs/2002ApJ...568..335M} {568, 335}

\bibitem[\protect\citeauthoryear{Marley, Saumon, Morley  \& Fortney}{Marley
  et~al.}{2018}]{sonora18}
Marley M.,  Saumon D.,  Morley C.,   Fortney J.,  2018, {Sonora 2018:
  Cloud-free, solar composition, solar C/O substellar atmosphere models and
  spectra}, \mn@doi{10.5281/zenodo.1309035}, \url
  {https://doi.org/10.5281/zenodo.1309035}

\bibitem[\protect\citeauthoryear{Marley et~al.,}{Marley
  et~al.}{2021}]{marley21}
Marley M.~S.,  et~al., 2021, \mn@doi [\apj] {10.3847/1538-4357/ac141d}, 920, 85

\bibitem[\protect\citeauthoryear{{Maxted}}{{Maxted}}{2016}]{ellc}
{Maxted} P.~F.~L.,  2016, \mn@doi [\aap] {10.1051/0004-6361/201628579}, \href
  {https://ui.adsabs.harvard.edu/abs/2016A&A...591A.111M} {591, A111}

\bibitem[\protect\citeauthoryear{{Mayorga}, {Robinson}, {Marley}, {May}  \&
  {Stevenson}}{{Mayorga} et~al.}{2021}]{mayorga21}
{Mayorga} L.~C.,  {Robinson} T.~D.,  {Marley} M.~S.,  {May} E.~M.,
  {Stevenson} K.~B.,  2021, \mn@doi [\apj] {10.3847/1538-4357/abff50}, \href
  {https://ui.adsabs.harvard.edu/abs/2021ApJ...915...41M} {915, 41}

\bibitem[\protect\citeauthoryear{{Mireles} et~al.,}{{Mireles}
  et~al.}{2020}]{toi694}
{Mireles} I.,  et~al., 2020, \mn@doi [\aj] {10.3847/1538-3881/aba526}, \href
  {https://ui.adsabs.harvard.edu/abs/2020AJ....160..133M} {160, 133}

\bibitem[\protect\citeauthoryear{{Mordasini}, {Alibert}  \& {Benz}}{{Mordasini}
  et~al.}{2009}]{mordasini09}
{Mordasini} C.,  {Alibert} Y.,   {Benz} W.,  2009, \mn@doi [A\&A]
  {10.1051/0004-6361/200810301}, \href
  {https://ui.adsabs.harvard.edu/abs/2009A&A...501.1139M} {501, 1139}

\bibitem[\protect\citeauthoryear{{Moutou} et~al.,}{{Moutou}
  et~al.}{2013}]{koi415b}
{Moutou} C.,  et~al., 2013, \mn@doi [\aap] {10.1051/0004-6361/201322201}, \href
  {https://ui.adsabs.harvard.edu/abs/2013A&A...558L...6M} {558, L6}

\bibitem[\protect\citeauthoryear{{Nakajima}, {Oppenheimer}, {Kulkarni},
  {Golimowski}, {Matthews}  \& {Durrance}}{{Nakajima} et~al.}{1995}]{gliese}
{Nakajima} T.,  {Oppenheimer} B.~R.,  {Kulkarni} S.~R.,  {Golimowski} D.~A.,
  {Matthews} K.,   {Durrance} S.~T.,  1995, \mn@doi [\nat] {10.1038/378463a0},
  \href {https://ui.adsabs.harvard.edu/abs/1995Natur.378..463N} {378, 463}

\bibitem[\protect\citeauthoryear{{Nefs} et~al.,}{{Nefs} et~al.}{2013}]{wts19g}
{Nefs} S.~V.,  et~al., 2013, \mn@doi [\mnras] {10.1093/mnras/stt405}, \href
  {https://ui.adsabs.harvard.edu/abs/2013MNRAS.431.3240N} {431, 3240}

\bibitem[\protect\citeauthoryear{{Nowak} et~al.,}{{Nowak}
  et~al.}{2017}]{cww89ab1}
{Nowak} G.,  et~al., 2017, \mn@doi [\aj] {10.3847/1538-3881/aa5cb6}, \href
  {https://ui.adsabs.harvard.edu/abs/2017AJ....153..131N} {153, 131}

\bibitem[\protect\citeauthoryear{{Ofir}, {Gandolfi}, {Buchhave}, {Lacy},
  {Hatzes}  \& {Fridlund}}{{Ofir} et~al.}{2012}]{kic1571}
{Ofir} A.,  {Gandolfi} D.,  {Buchhave} L.,  {Lacy} C.~H.~S.,  {Hatzes} A.~P.,
  {Fridlund} M.,  2012, \mn@doi [\mnras] {10.1111/j.1745-3933.2011.01191.x},
  \href {https://ui.adsabs.harvard.edu/abs/2012MNRAS.423L...1O} {423, L1}

\bibitem[\protect\citeauthoryear{{Palle} et~al.,}{{Palle}
  et~al.}{2021}]{toi263b2}
{Palle} E.,  et~al., 2021, \mn@doi [\aap] {10.1051/0004-6361/202039937}, \href
  {https://ui.adsabs.harvard.edu/abs/2021A&A...650A..55P} {650, A55}

\bibitem[\protect\citeauthoryear{Parsons et~al.,}{Parsons
  et~al.}{2018}]{parsons18}
Parsons S.~G.,  et~al., 2018, \mn@doi [\mnras] {10.1093/mnras/sty2345}, 481,
  1083

\bibitem[\protect\citeauthoryear{Parviainen \& Aigrain}{Parviainen \&
  Aigrain}{2015}]{ldtk}
Parviainen H.,  Aigrain S.,  2015, \mn@doi [MNRAS] {10.1093/mnras/stv1857},
  453, 3821

\bibitem[\protect\citeauthoryear{{Parviainen} et~al.,}{{Parviainen}
  et~al.}{2020}]{toi263b1}
{Parviainen} H.,  et~al., 2020, \mn@doi [\aap] {10.1051/0004-6361/201935958},
  \href {https://ui.adsabs.harvard.edu/abs/2020A&A...633A..28P} {633, A28}

\bibitem[\protect\citeauthoryear{{Pearson} et~al.,}{{Pearson}
  et~al.}{2022}]{pearson22}
{Pearson} K.~A.,  et~al., 2022, \mn@doi [\aj] {10.3847/1538-3881/ac8dee}, \href
  {https://ui.adsabs.harvard.edu/abs/2022AJ....164..178P} {164, 178}

\bibitem[\protect\citeauthoryear{{Pecaut} \& {Mamajek}}{{Pecaut} \&
  {Mamajek}}{2013}]{mamajek}
{Pecaut} M.~J.,  {Mamajek} E.~E.,  2013, \mn@doi [\apjs]
  {10.1088/0067-0049/208/1/9}, \href
  {https://ui.adsabs.harvard.edu/abs/2013ApJS..208....9P} {208, 9}

\bibitem[\protect\citeauthoryear{{Persson} et~al.,}{{Persson}
  et~al.}{2019}]{epic212036875b1}
{Persson} C.~M.,  et~al., 2019, \mn@doi [\aap] {10.1051/0004-6361/201935505},
  \href {https://ui.adsabs.harvard.edu/abs/2019A&A...628A..64P} {628, A64}

\bibitem[\protect\citeauthoryear{{Pont}, {Melo}, {Bouchy}, {Udry}, {Queloz},
  {Mayor}  \& {Santos}}{{Pont} et~al.}{2005a}]{ogle122b}
{Pont} F.,  {Melo} C.~H.~F.,  {Bouchy} F.,  {Udry} S.,  {Queloz} D.,  {Mayor}
  M.,   {Santos} N.~C.,  2005a, \mn@doi [\aap] {10.1051/0004-6361:200500025},
  \href {https://ui.adsabs.harvard.edu/abs/2005A&A...433L..21P} {433, L21}

\bibitem[\protect\citeauthoryear{{Pont}, {Bouchy}, {Melo}, {Santos}, {Mayor},
  {Queloz}  \& {Udry}}{{Pont} et~al.}{2005b}]{ogle106b}
{Pont} F.,  {Bouchy} F.,  {Melo} C.,  {Santos} N.~C.,  {Mayor} M.,  {Queloz}
  D.,   {Udry} S.,  2005b, \mn@doi [\aap] {10.1051/0004-6361:20052771}, \href
  {https://ui.adsabs.harvard.edu/abs/2005A&A...438.1123P} {438, 1123}

\bibitem[\protect\citeauthoryear{{Pont} et~al.,}{{Pont}
  et~al.}{2006}]{ogle123b}
{Pont} F.,  et~al., 2006, \mn@doi [\aap] {10.1051/0004-6361:20053692}, \href
  {https://ui.adsabs.harvard.edu/abs/2006A&A...447.1035P} {447, 1035}

\bibitem[\protect\citeauthoryear{Prantzos et~al.,}{Prantzos
  et~al.}{2023}]{discages}
Prantzos N.,  et~al., 2023, \mn@doi [Monthly Notices of the Royal Astronomical
  Society] {10.1093/mnras/stad1551}, 523, 2126

\bibitem[\protect\citeauthoryear{{Psaridi} et~al.,}{{Psaridi}
  et~al.}{2022}]{psaridi22}
{Psaridi} A.,  et~al., 2022, arXiv e-prints, \href
  {https://ui.adsabs.harvard.edu/abs/2022arXiv220510854P} {p. arXiv:2205.10854}

\bibitem[\protect\citeauthoryear{{Rebolo}, {Zapatero Osorio}  \&
  {Mart{\'\i}n}}{{Rebolo} et~al.}{1995}]{teide}
{Rebolo} R.,  {Zapatero Osorio} M.~R.,   {Mart{\'\i}n} E.~L.,  1995, \mn@doi
  [\nat] {10.1038/377129a0}, \href
  {https://ui.adsabs.harvard.edu/abs/1995Natur.377..129R} {377, 129}

\bibitem[\protect\citeauthoryear{Ricker et~al.,}{Ricker et~al.}{2014}]{tess}
Ricker G.~R.,  et~al., 2014, \mn@doi [Journal of Astronomical Telescopes,
  Instruments, and Systems] {10.1117/1.JATIS.1.1.014003}, 1, 1

\bibitem[\protect\citeauthoryear{{Robinson} \& {Marley}}{{Robinson} \&
  {Marley}}{2014}]{robinson14}
{Robinson} T.~D.,  {Marley} M.~S.,  2014, \mn@doi [\apj]
  {10.1088/0004-637X/785/2/158}, \href
  {https://ui.adsabs.harvard.edu/abs/2014ApJ...785..158R} {785, 158}

\bibitem[\protect\citeauthoryear{{Schaffenroth} et~al.,}{{Schaffenroth}
  et~al.}{2021}]{schaffenroth}
{Schaffenroth} V.,  et~al., 2021, \mn@doi [\mnras] {10.1093/mnras/staa3661},
  \href {https://ui.adsabs.harvard.edu/abs/2021MNRAS.501.3847S} {501, 3847}

\bibitem[\protect\citeauthoryear{Sebastian et~al.,}{Sebastian
  et~al.}{2022}]{corot34b}
Sebastian D.,  et~al., 2022, \mn@doi [Monthly Notices of the Royal Astronomical
  Society] {10.1093/mnras/stac2131}, 516, 636

\bibitem[\protect\citeauthoryear{{Shporer} et~al.,}{{Shporer}
  et~al.}{2017}]{k276}
{Shporer} A.,  et~al., 2017, \mn@doi [\apjl] {10.3847/2041-8213/aa8bff}, \href
  {https://ui.adsabs.harvard.edu/abs/2017ApJ...847L..18S} {847, L18}

\bibitem[\protect\citeauthoryear{{Siverd} et~al.,}{{Siverd}
  et~al.}{2012}]{kelt1b}
{Siverd} R.~J.,  et~al., 2012, \mn@doi [\apj] {10.1088/0004-637X/761/2/123},
  \href {https://ui.adsabs.harvard.edu/abs/2012ApJ...761..123S} {761, 123}

\bibitem[\protect\citeauthoryear{{Skrutskie} et~al.,}{{Skrutskie}
  et~al.}{2006}]{2mass}
{Skrutskie} M.~F.,  et~al., 2006, \mn@doi [\aj] {10.1086/498708}, \href
  {https://ui.adsabs.harvard.edu/abs/2006AJ....131.1163S} {131, 1163}

\bibitem[\protect\citeauthoryear{{Smith} et~al.,}{{Smith}
  et~al.}{2012}]{smith12}
{Smith} J.~C.,  et~al., 2012, \mn@doi [\pasp] {10.1086/667697}, \href
  {http://adsabs.harvard.edu/abs/2012PASP..124.1000S} {124, 1000}

\bibitem[\protect\citeauthoryear{{Speagle}}{{Speagle}}{2020}]{dynesty}
{Speagle} J.~S.,  2020, \mn@doi [\mnras] {10.1093/mnras/staa278}, \href
  {https://ui.adsabs.harvard.edu/abs/2020MNRAS.493.3132S} {493, 3132}

\bibitem[\protect\citeauthoryear{{Stassun}, {Mathieu}  \& {Valenti}}{{Stassun}
  et~al.}{2006}]{Stassun06}
{Stassun} K.~G.,  {Mathieu} R.~D.,   {Valenti} J.~A.,  2006, \mn@doi [\nat]
  {10.1038/nature04570}, \href
  {https://ui.adsabs.harvard.edu/abs/2006Natur.440..311S} {440, 311}

\bibitem[\protect\citeauthoryear{{Stumpe} et~al.,}{{Stumpe}
  et~al.}{2012}]{Stumpe2012}
{Stumpe} M.~C.,  et~al., 2012, \mn@doi [\pasp] {10.1086/667698}, \href
  {https://ui.adsabs.harvard.edu/abs/2012PASP..124..985S} {124, 985}

\bibitem[\protect\citeauthoryear{{Stumpe}, {Smith}, {Catanzarite}, {Van Cleve},
  {Jenkins}, {Twicken}  \& {Girouard}}{{Stumpe} et~al.}{2014}]{Stumpe2014}
{Stumpe} M.~C.,  {Smith} J.~C.,  {Catanzarite} J.~H.,  {Van Cleve} J.~E.,
  {Jenkins} J.~M.,  {Twicken} J.~D.,   {Girouard} F.~R.,  2014, \mn@doi [\pasp]
  {10.1086/674989}, \href {http://adsabs.harvard.edu/abs/2014PASP..126..100S}
  {126, 100}

\bibitem[\protect\citeauthoryear{{Tal-Or} et~al.,}{{Tal-Or}
  et~al.}{2013}]{corot1011}
{Tal-Or} L.,  et~al., 2013, \mn@doi [\aap] {10.1051/0004-6361/201220862}, \href
  {https://ui.adsabs.harvard.edu/abs/2013A&A...553A..30T} {553, A30}

\bibitem[\protect\citeauthoryear{{Tokovinin}}{{Tokovinin}}{2018}]{Tokovinin}
{Tokovinin} A.,  2018, \mn@doi [\pasp] {10.1088/1538-3873/aaa7d9}, \href
  {https://ui.adsabs.harvard.edu/abs/2018PASP..130c5002T} {130, 035002}

\bibitem[\protect\citeauthoryear{Tokovinin \& Kiyaeva}{Tokovinin \&
  Kiyaeva}{2015}]{binaryavecc}
Tokovinin A.,  Kiyaeva O.,  2015, \mn@doi [Monthly Notices of the Royal
  Astronomical Society] {10.1093/mnras/stv2825}, 456, 2070

\bibitem[\protect\citeauthoryear{{Triaud} et~al.,}{{Triaud}
  et~al.}{2013}]{wasp30b2}
{Triaud} A.~H.~M.~J.,  et~al., 2013, \mn@doi [\aap]
  {10.1051/0004-6361/201219643}, \href
  {https://ui.adsabs.harvard.edu/abs/2013A&A...549A..18T} {549, A18}

\bibitem[\protect\citeauthoryear{{Triaud} et~al.,}{{Triaud}
  et~al.}{2017}]{triaud17}
{Triaud} A. H.~M.~J.,  et~al., 2017, \mn@doi [\aap]
  {10.1051/0004-6361/201730993}, \href
  {https://ui.adsabs.harvard.edu/abs/2017A&A...608A.129T} {608, A129}

\bibitem[\protect\citeauthoryear{Tsai, Steinrueck, Parmentier, Lewis  \&
  Pierrehumbert}{Tsai et~al.}{2023}]{tsai23}
Tsai S.-M.,  Steinrueck M.,  Parmentier V.,  Lewis N.,   Pierrehumbert R.,
  2023, \mn@doi [Monthly Notices of the Royal Astronomical Society]
  {10.1093/mnras/stad214}, 520, 3867

\bibitem[\protect\citeauthoryear{{Twicken} et~al.,}{{Twicken}
  et~al.}{2018}]{Twicken}
{Twicken} J.~D.,  et~al., 2018, \mn@doi [\pasp] {10.1088/1538-3873/aab694},
  \href {http://adsabs.harvard.edu/abs/2018PASP..130f4502T} {130, 064502}

\bibitem[\protect\citeauthoryear{{Vowell} et~al.,}{{Vowell}
  et~al.}{2023}]{hip33609b}
{Vowell} N.,  et~al., 2023, \mn@doi [\aj] {10.3847/1538-3881/acd197}, \href
  {https://ui.adsabs.harvard.edu/abs/2023AJ....165..268V} {165, 268}

\bibitem[\protect\citeauthoryear{{Wheatley} et~al.,}{{Wheatley}
  et~al.}{2018}]{NGTS}
{Wheatley} P.~J.,  et~al., 2018, \mn@doi [\mnras] {10.1093/mnras/stx2836},
  \href {https://ui.adsabs.harvard.edu/abs/2018MNRAS.475.4476W} {475, 4476}

\bibitem[\protect\citeauthoryear{{Whitworth}}{{Whitworth}}{2018}]{whitworth18}
{Whitworth} A.,  2018, arXiv e-prints, \href
  {https://ui.adsabs.harvard.edu/abs/2018arXiv181106833W} {p. arXiv:1811.06833}

\bibitem[\protect\citeauthoryear{{Wittenmyer}, {Endl}, {Cochran},
  {Ram{\'\i}rez}, {Reffert}, {MacQueen}  \& {Shetrone}}{{Wittenmyer}
  et~al.}{2009}]{HD91669B}
{Wittenmyer} R.~A.,  {Endl} M.,  {Cochran} W.~D.,  {Ram{\'\i}rez} I.,
  {Reffert} S.,  {MacQueen} P.~J.,   {Shetrone} M.,  2009, \mn@doi [\aj]
  {10.1088/0004-6256/137/3/3529}, \href
  {https://ui.adsabs.harvard.edu/abs/2009AJ....137.3529W} {137, 3529}

\bibitem[\protect\citeauthoryear{{York} et~al.,}{{York}
  et~al.}{2000}]{sdsstech}
{York} D.~G.,  et~al., 2000, \mn@doi [\aj] {10.1086/301513}, \href
  {https://ui.adsabs.harvard.edu/abs/2000AJ....120.1579Y} {120, 1579}

\bibitem[\protect\citeauthoryear{{Zahn} \& {Bouchet}}{{Zahn} \&
  {Bouchet}}{1989}]{zahn89}
{Zahn} J.~P.,  {Bouchet} L.,  1989, \aap, \href
  {https://ui.adsabs.harvard.edu/abs/1989A&A...223..112Z} {223, 112}

\bibitem[\protect\citeauthoryear{{Zhou} et~al.,}{{Zhou}
  et~al.}{2014}]{hats016b}
{Zhou} G.,  et~al., 2014, \mn@doi [\mnras] {10.1093/mnras/stt2100}, \href
  {https://ui.adsabs.harvard.edu/abs/2014MNRAS.437.2831Z} {437, 2831}

\bibitem[\protect\citeauthoryear{{Zhou} et~al.,}{{Zhou} et~al.}{2019}]{hats70b}
{Zhou} G.,  et~al., 2019, \mn@doi [\aj] {10.3847/1538-3881/aaf1bb}, \href
  {https://ui.adsabs.harvard.edu/abs/2019AJ....157...31Z} {157, 31}

\bibitem[\protect\citeauthoryear{{Ziegler}, {Tokovinin}, {Brice{\~n}o}, {Mang},
  {Law}  \& {Mann}}{{Ziegler} et~al.}{2020}]{ziegler}
{Ziegler} C.,  {Tokovinin} A.,  {Brice{\~n}o} C.,  {Mang} J.,  {Law} N.,
  {Mann} A.~W.,  2020, \mn@doi [\aj] {10.3847/1538-3881/ab55e9}, \href
  {https://ui.adsabs.harvard.edu/abs/2020AJ....159...19Z} {159, 19}

\bibitem[\protect\citeauthoryear{{{\v{S}}ubjak} et~al.,}{{{\v{S}}ubjak}
  et~al.}{2020}]{toi503b}
{{\v{S}}ubjak} J.,  et~al., 2020, \mn@doi [\aj] {10.3847/1538-3881/ab7245},
  \href {https://ui.adsabs.harvard.edu/abs/2020AJ....159..151S} {159, 151}

\bibitem[\protect\citeauthoryear{{von Boetticher} et~al.,}{{von Boetticher}
  et~al.}{2017}]{j055557ab}
{von Boetticher} A.,  et~al., 2017, \mn@doi [\aap]
  {10.1051/0004-6361/201731107}, \href
  {https://ui.adsabs.harvard.edu/abs/2017A&A...604L...6V} {604, L6}

\bibitem[\protect\citeauthoryear{{von Boetticher} et~al.,}{{von Boetticher}
  et~al.}{2019}]{j0954}
{von Boetticher} A.,  et~al., 2019, \mn@doi [\aap]
  {10.1051/0004-6361/201834539}, \href
  {https://ui.adsabs.harvard.edu/abs/2019A&A...625A.150V} {625, A150}

\makeatother
\end{thebibliography}


\appendix
\section{Extra Data, figures and tables}\label{appendix:section}
\subsection{Full Data Tables}\label{appendix:tables}

\begin{table*}
\caption{Parameters of all objects used within the population analysis. RIK-72b \citep{rik72b} and the binary system discovered by \citet{Stassun06} are not included due to their high levels of inflation. Table has been updated from \citet{henderson24} and \citet{grieves21}. Sources: [1]: \citet{toi4603b}, [2]:\citet{hats70b}, [3]: \citet{toi1278b}, [4]: \citet{gpx1b}, [5]: \citet{kepler39b}, [6]: \citet{corot3b}, [7]: \citet{kelt1b}, [8]: \citet{beatty14_1b}, [9]: \citet{nltt}, [10]: \citet{NLTTrad}, [11]: \citet{wasp128b}, [12]: \citet{cww89ab1}, [13]: \citet{carmichael19}, [14]: \citet{koi205b}, [15]:  \citet{carmichael20}, [16]: \citet{epic212036875b1}, [17]: \citet{toi503b}, [18]: \citet{carmichael21}, [19]: \citet{ad3116b}, [20]: \citet{corot33b}, [21]: \citet{toi263b1}, [22]: \citet{toi263b2}, [23]: \citet{koi415b}, [24]: \citet{wasp30b2}, [25]: \citet{lhs6343}, [26]: \citet{corot15b}, [27]: \citet{carmichael22}, [28]: \citet{psaridi22}, [29]: \citet{henderson24}, [30]: \citet{epic201702477b}, [31]: \citet{hip33609b}, [32]: \citet{lp261}, [33]: \citet{19b}, [34]: \citet{lin23}, [35]: \citet{corot34b}, [36]: \citet{ngts7ab}, [37]: \citet{toi5375b}, [38]: \citet{grieves21}, [39]: \citet{koi189b}, [40]: \citet{j055557ab}, [41]: \citet{j0954}, [42]: \citet{ogle123b}, [43]: \citet{toi694}, [44]: \citet{tic320687387b},[45]:\citet{ogle122b}, [46]: \citet{k276}, [47]: \citet{corot1011}, [48]: \citet{j2343}, [49]: \citet{hats016b}, [50]: \citet{ogle106b}, [51]: \citet{hat205}, [52]: \citet{tic2310}, [53]: \citet{kic1571}, [54]: \citet{wts19g}. Sources with $\ast$ have updated radii from \citet{imporvedradii}.}            
\label{tab:full_obj_lists}      
\centering   
\begin{tabular}{>{\bfseries}l l l l l l l l l l}
\hline\hline                        
Object & \textbf{P [d]} & \textbf{M$_{2}$ [\mjup]} & \textbf{R$_{2}$ [\rjup]} & \textbf{\teff [K] } & \textbf{M$_{1}$ [\msun]} & \textbf{R$_{1}$ [\rsun]} & \textbf{ecc} & \textbf{[Fe/H]} & \textbf{Source}\\
\hline
TOI-4603b&$7.25$&$12.9^{+0.58}_{-0.57}$&$1.04\pm0.04$&$6264^{+95}_{-94}$&$1.77\pm0.06$&$2.74\pm0.05$&$0.325\pm0.020$&$0.34\pm0.04$& [1]\\
HATS-70b&$1.89$&$12.9^{+1.8}_{-1.6}$&$1.38^{+0.08}_{-0.07}$&$7930^{+630}_{-820}$&$1.78\pm0.12$&$1.88^{+0.06}_{-0.07}$&<$0.18$&$0.04^{+0.10}_{-0.11}$& [2]\\
TOI-1278b&$14.48$&$18.5\pm0.5$&$1.09^{+0.24}_{-0.20}$&$3799\pm42$&$0.55\pm0.02$&$0.57\pm0.01$&$0.013\pm0.004$&$-0.01\pm0.28$& [3]\\
GPX-1b&$1.74$&$19.7\pm1.6$&$1.47\pm0.10$&$7000\pm200$&$1.68\pm0.10$&$1.56\pm0.10$&$0$ (fixed)&$0.35\pm0.10$& [4]\\
Kepler-39b&$21.09$&$20.1^{+1.3}_{-1.2}$&$1.07\pm0.03$$\ast$&$6350\pm100$&$1.29^{+0.06}_{-0.07}$&$1.40\pm0.10$&$0.112\pm0.057$&$0.10\pm0.14$& [5]\\
CoRoT-3b&$4.26$&$21.7\pm1.0$&$1.08\pm0.05$$\ast$&$6740\pm140$&$1.37\pm0.09$&$1.56\pm0.09$&$0$ (fixed)&$-0.02\pm0.06$& [6]\\
KELT-1b&$1.22$&$27.4\pm0.9$&$1.13\pm0.03$$\ast$&$6516\pm49$&$1.34\pm0.06$&$1.47^{+0.05}_{-0.04}$&$0.010^{+0.010}_{-0.007}$&$0.05\pm0.08$& [7][8]\\
NLTT41135b&$2.89$&$33.7^{+2.8}_{-2.6}$&$1.13^{+0.27}_{-0.17}$&$3230\pm130$&$0.19^{+0.03}_{-0.02}$&$0.21^{+0.02}_{-0.01}$&<$0.02$&$0$ (fixed)& [9][10] \\
WASP-128b&$2.21$&$37.2^{+0.8}_{-0.9}$&$0.96\pm0.02$$\ast$&$5950\pm50$&$1.16\pm0.04$&$1.15\pm0.02$&<$0.007$&$0.01\pm0.12$& [11]\\
CWW89Ab&$5.29$&$39.2\pm1.1$&$0.94\pm0.02$&$5755\pm49$&$1.10\pm0.05$&$1.03\pm0.02$&$0.189\pm0.002$&$0.20\pm0.09$& [12][13]\\
KOI-205b&$11.72$&$39.9\pm1.0$&$0.87\pm0.02$$\ast$&$5237\pm60$&$0.93\pm0.03$&$0.84\pm0.02$&<$0.031$&$0.14\pm0.12$& [14]\\
TOI-1406b&$10.57$&$46.0^{+2.6}_{-2.7}$&$0.86\pm0.03$&$6290\pm100$&$1.18^{+0.08}_{-0.09}$&$1.35\pm0.03$&$0.026^{+0.013}_{-0.010}$&$-0.08\pm0.09$& [15]\\
EPIC212036875b&$ 5.17$&$52.3\pm1.9$&$0.87\pm0.02$&$6238^{+59}_{-60}$&$1.29^{+0.07}_{-0.06}$&$1.50\pm0.03$&$0.132\pm0.004$&$0.01\pm0.10$&[13][16] \\
TOI-503b&$3.68$&$53.7\pm1.2$&$1.34^{+0.26}_{-0.15}$&$7650^{+140}_{-160}$&$1.80\pm0.06$&$1.70^{+0.05}_{-0.04}$&$0$ (fixed)&$0.30^{+0.08}_{-0.09}$& [17]\\
TOI-852b&$4.95$&$53.7^{+1.4}_{-1.3}$&$0.83\pm0.04$&$5768^{+84}_{-81}$&$1.32^{+0.05}_{-0.04}$&$1.71\pm0.04$&$0.004^{+0.004}_{-0.003}$&$0.33\pm0.09$& [18]\\
AD3116b&$1.98$&$54.2\pm4.3$&$0.95\pm0.07$$\ast$&$3184\pm29$&$0.28\pm0.02$&$0.29\pm0.08$&$0.146^{+0.024}_{-0.016}$&$0$ (fixed)& [19] \\
CoRoT-33b&$5.82$&$59.0^{+1.8}_{-1.7}$&$1.10\pm0.53$&$5225\pm80$&$0.86\pm0.04$&$0.94^{+0.14}_{-0.08}$&$0.070\pm0.002$&$0.44\pm0.10$& [20] \\
TOI-811b&$25.17$&$59.9^{+13.0}_{-8.6}$&$1.26\pm0.06$&$6107\pm77$&$1.32^{+0.05}_{-0.07}$&$1.27^{+0.06}_{-0.09}$&$0.509\pm0.075$&$0.40^{+0.07}_{-0.09}$& [18]\\
TOI-263b&$0.56$&$61.6\pm4.0$&$0.91\pm0.07$&$3471\pm33$&$0.44\pm0.04$&$0.44\pm0.03$&$0.017^{+0.009}_{-0.010}$&$0.00\pm0.10$& [21][22]\\
KOI-415b&$166.79$&$62.1\pm2.7$&$0.86\pm0.03$$\ast$&$5810\pm80$&$0.94\pm0.06$&$1.25^{+0.15}_{-0.10}$&$0.698\pm0.002$&$-0.24\pm0.11$& [23]\\
WASP-30b&$4.16$&$62.5\pm1.2$&$0.96\pm0.03$$\ast$&$6202^{+42}_{-51}$&$1.25^{+0.03}_{-0.04}$&$1.39\pm0.03$&<$0.004$&$0.08^{+0.07}_{-0.05}$& [24]\\
LHS6343c&$12.71$&$62.7\pm2.4$&$0.83\pm0.02$&$3130\pm20$&$0.37\pm0.01$&$0.38\pm0.01$&$0.056\pm0.032$&$0.04\pm0.08$& [25]\\
CoRoT-15b&$3.06$&$63.3\pm4.1$&$0.94\pm0.12$$\ast$&$6350\pm200$&$1.32\pm0.12$&$1.46^{+0.31}_{-0.14}$&$0$ (fixed)&$0.10\pm0.20$& [26]\\
TOI-569b&$6.56$&$64.1^{+1.9}_{-1.4}$&$0.75\pm0.02$&$5768^{+110}_{-92}$&$1.21\pm0.05$&$1.48\pm0.03$&$0.002^{+0.002}_{-0.001}$&$0.29^{+0.09}_{-0.08}$& [15]\\
TOI-2119b&$7.20$&$64.4^{+2.3}_{-2.2}$&$1.08\pm0.03$&$3621^{+48}_{-46}$&$0.53\pm0.02$&$0.50\pm0.02$&$0.337^{+0.002}_{-0.001}$&$0.06\pm0.08$& [27]\\
TOI-1982b&$17.17$&$65.9^{+2.8}_{-2.7}$&$1.08\pm0.04$&$6325\pm110$&$1.41\pm0.08$&$1.51\pm0.05$&$0.272\pm0.014$&$-0.10\pm0.09$& [28]\\
NGTS-28Ab&$1.25$&$67.7^{+5.4}_{-4.9}$&$0.97\pm0.05$&$3626^{+47}_{-44}$&$0.56\pm0.02$&$0.59\pm0.03$&$0.041^{+0.005}_{-0.008}$&$-0.14^{+0.16}_{-0.17}$& [29]\\
EPIC201702477b&$40.74$&$66.9\pm1.7$&$0.83\pm0.04$$\ast$&$5517\pm70$&$0.87\pm0.03$&$0.90\pm0.06$&$0.228\pm0.003$&$-0.16\pm0.05$& [30]\\
TOI-629b&$8.72$&$67.0\pm3.0$&$1.11\pm0.05$&$9100\pm200$&$2.16\pm0.13$&$2.37\pm0.11$&$0.298\pm0.008$&$0.10\pm0.15$& [28]\\
TOI-2543b&$7.54$&$67.6\pm3.5$&$0.95\pm0.09$&$6060\pm82$&$1.29\pm0.08$&$1.86\pm0.15$&$0.009^{+0.003}_{-0.002}$&$-0.28\pm0.10$& [28]\\
HIP-33609b&$39.47$&$68.0^{+7.4}_{-7.1}$&$1.58\pm0.07$&$10400^{+800}_{-660}$&$2.38\pm0.10$&$1.86^{+0.09}_{-0.08}$&$0.560^{+0.029}_{-0.031}$&$-0.01^{+0.19}_{-0.20}$& [31]\\
LP261-75b&$1.88$&$68.1\pm2.1$&$0.90\pm0.01$&$3100\pm50$&$0.30\pm0.02$&$0.31\pm0.00$&<$0.007$&-& [32]\\
NGTS-19b&$17.84$&$69.5^{+5.7}_{-5.4}$&$1.03^{+0.06}_{-0.05}$&$4716^{+39}_{-28}$&$0.81\pm0.04$&$0.90\pm0.04$&$0.377\pm0.006$&$0.11\pm0.07$& [33]\\
TOI-2336b&$7.71$&$69.9\pm2.3$&$1.05\pm0.04$&$6433\pm84$&$1.40\pm0.07$&$1.82\pm0.06$&$0.010^{+0.006}_{-0.005}$&$0.09\pm0.11$& [34]\\
CoRoT-34b&$2.12$&$71.4^{+8.9}_{-8.6}$&$1.09^{+0.17}_{-0.16}$&$7820\pm160$&$1.66^{+0.08}_{-0.15}$&$1.85^{+0.29}_{-0.25}$&$0$ (fixed)&$-0.20\pm0.20$& [35]\\
\textcolor{red}{TOI-2490b}&60.33&$73.6\pm2.4$&$1.00\pm0.02$&$5558\pm80$&$1.00^{+0.03}_{-0.02}$&$1.11\pm0.012$&$0.780\pm0.000$&$0.32\pm0.05$&This work\\
NGTS-7Ab&$0.68$&$75.5^{+3.0}_{-13.7}$&$1.38^{+0.13}_{-0.14}$&$3359^{+106}_{-89}$&$0.48^{+0.03}_{-0.12}$&$0.61\pm0.06$&$0$ (fixed)&$0$ (fixed)& [36]\\
TOI-5375b&$1.72$&$77\pm8.0$&$0.99\pm0.16$&$3885\pm25$&$0.64\pm0.10$&$0.62\pm0.10$&$0$ (fixed)&$0.26\pm0.04$& [37]\\
TOI-148b&$4.87$&$77.1^{+5.8}_{-4.6}$&$0.81^{+0.05}_{-0.06}$&$5990\pm140$&$0.97^{+0.12}_{-0.09}$&$1.20\pm0.07$&$0.005^{+0.006}_{-0.004}$&$-0.24\pm0.25$& [38]\\
TOI-2521b&$5.56$&$77.5\pm3.3$&$1.01\pm0.04$&$5625\pm74$&$0.95\pm0.06$&$1.74\pm0.06$&<$0.035$&$-0.28\pm0.13$& [34]\\
KOI-189b&$30.36$&$78.0\pm3.4$&$0.99\pm0.02$$\ast$&$4952\pm40$&$0.76\pm0.05$&$0.73\pm0.02$&$0.275\pm0.004$&$-0.12\pm0.10$& [39]\\
TOI-587b&$8.04$&$81.1^{+7.1}_{-7.0}$&$1.32^{+0.07}_{-0.06}$&$9800\pm200$&$2.33\pm0.12$&$2.01\pm0.09$&$0.051^{+0.049}_{-0.036}$&$0.08^{+0.11}_{-0.12}$& [38]\\
TOI-746b&$10.98$&$82.2^{+4.9}_{-4.4}$&$0.95^{+0.09}_{-0.06}$&$5690\pm140$&$0.94^{+0.09}_{-0.08}$&$0.97^{+0.04}_{-0.03}$&$0.199\pm0.003$&$-0.02\pm0.23$& [38]\\
EBLM-J0555-57Ab&$7.76$&$87.9\pm4.0$&$0.82^{+0.13}_{-0.06}$&$6368\pm124$&$1.18\pm0.08$&$1.00^{+0.14}_{-0.07}$&$0.090\pm0.004$&$-0.04\pm0.14$& [40][41]\\
TOI-681b&$15.78$&$88.7^{+2.5}_{-2.3}$&$1.52^{+0.25}_{-0.15}$&$7440^{+150}_{-140}$&$1.54^{+0.06}_{-0.05}$&$1.47\pm0.04$&$0.093^{+0.022}_{-0.019}$&$-0.08\pm0.05$& [38]\\
OGLE-TR-123b&$1.80$&$89.0\pm11.5$&$1.29\pm0.09$&$6700\pm300$&$1.29\pm0.26$&$1.55\pm0.10$&$0$ (fixed)&-& [42]\\
TOI-694b&$48.05$&$89.0\pm5.3$&$1.11\pm0.02$&$5496^{+87}_{-81}$&$0.97^{+0.05}_{-0.04}$&$1.00\pm0.01$&$0.519\pm0.001$&$0.21\pm0.08$& [43]\\
TOI-1608b&$2.47$&$90.7\pm3.7$&$1.21\pm0.06$&$6028\pm82$&$1.31\pm0.07$&$2.16\pm0.08$&$0.041^{+0.024}_{-0.019}$&$0.09\pm0.12$& [34]\\
KOI-607b&$5.89$&$95.1^{+3.3}_{-3.4}$&$1.09^{+0.09}_{-0.06}$&$5418^{+87}_{-85}$&$0.99\pm0.05$&$0.92\pm0.03$&$0.395\pm0.009$&$0.38^{+0.08}_{-0.09}$& [13]\\
J1219-39b&$6.76$&$95.4^{+1.9}_{-2.5}$&$1.14^{+0.07}_{-0.05}$&$5412^{+81}_{-65}$&$0.83\pm0.03$&$0.81^{+0.04}_{-0.02}$&$0.055\pm0.000$&$-0.21^{+0.07}_{-0.08}$& [24]\\
\multicolumn{10}{l}{}
\end{tabular}
\end{table*}

\begin{table*}
\centering
\contcaption{}
\begin{tabular}{>{\bfseries}l l l l l l l l l l}
TIC320687387 B&$29.77$&$96.2^{+1.9}_{-2.0}$&$1.14\pm0.02$&$5780\pm80$&$1.08\pm0.03$&$1.16\pm0.02$&$0.366\pm0.003$&$0.30\pm0.08$& [44]\\
OGLE-TR-122b&$7.27$&$96.4\pm9.4$&$1.17^{+0.23}_{-0.13}$&$5700\pm300$&$0.98\pm0.14$&$1.05^{+0.20}_{-0.09}$&$0.205\pm0.008$&$0.15\pm0.36$& [45]\\
TOI-1213b&$27.22$&$97.5^{+4.4}_{-4.2}$&$1.66^{+0.78}_{-0.55}$&$5590\pm150$&$0.99^{+0.07}_{-0.06}$&$0.99\pm0.04$&$0.498^{+0.003}_{-0.002}$&$0.25^{+0.13}_{-0.14}$& [38]\\
K2-76b&$11.99$&$98.7\pm2.0$&$0.89^{+0.05}_{-0.03}$&$5747^{+70}_{-64}$&$0.96\pm0.03$&$1.17^{+0.06}_{-0.03}$&$0.255\pm0.007$&$0.01\pm0.04$& [46]\\
CoRoT-101186644&$20.68$&$100.6\pm11.5$&$1.01^{+0.25}_{-0.06}$&$6090\pm200$&$1.20\pm0.20$&$1.07\pm0.07$&$0.402\pm0.006$&$0.20\pm0.20$& [47]\\
J2343+29Ab&$16.95$&$102.7\pm7.3$&$1.24\pm0.07$&$5150^{+90}_{-60}$&$0.86\pm0.10$&$0.85^{+0.05}_{-0.06}$&$0.161^{+0.002}_{-0.003}$&$0.07^{+0.01}_{-0.17}$& [48]\\
EBLM-J0954-23Ab&$7.57$&$102.8^{+5.9}_{-6.0}$&$0.98\pm0.17$&$6406\pm124$&$1.17\pm0.08$&$1.23\pm0.17$&$0.042\pm0.001$&$-0.01\pm0.14$& [41]\\
KOI-686b&$52.51$&$103.4\pm4.8$&$1.22\pm0.04$&$5834\pm100$&$0.98\pm0.07$&$1.04\pm0.03$&$0.556\pm0.004$&$-0.06\pm0.13$& [39]\\
TIC220568520b&$18.56$&$107.2\pm5.2$&$1.25\pm0.02$&$5589\pm81$&$1.03\pm0.04$&$1.01\pm0.01$&$0.096\pm0.003$&$0.26\pm0.07$& [43]\\
HATS550-016B&$2.05$&$115.2^{+5.2}_{-6.3}$&$1.43^{+0.03}_{-0.04}$&$6420\pm90$&$0.97^{+0.05}_{-0.06}$&$1.22^{+0.02}_{-0.03}$&$0.080\pm0.020$&$-0.60\pm0.06$& [49]\\
OGLE-TR-106b&$2.54$&$121.5\pm22.0$&$1.76\pm0.13$&-&-&$1.31\pm0.09$&$0.000\pm0.020$&-& [50]\\
EBLM-J1431-11Ab&$4.45$&$126.9^{+3.8}_{-3.9}$&$1.45^{+0.07}_{-0.05}$&$6161\pm124$&$1.20\pm0.06$&$1.11^{+0.04}_{-0.03}$&$0$ (fixed)&$0.15\pm0.14$& [41]\\
HAT-TR-205-013B&$2.23$&$129.9\pm10.5$&$1.63\pm0.06$&$6295\pm200$&$1.04\pm0.13$&$1.28\pm0.04$&$0.012\pm0.021$&-& [51]\\
TIC231005575b&$61.78$&$134.1\pm3.1$&$1.50\pm0.08$&$5500\pm85$&$1.05\pm0.04$&$0.99\pm0.05$&$0.298^{+0.001}_{-0.004}$&$-0.44\pm0.06$& [52]\\
HATS551-021B&$3.64$&$138.3^{+14.7}_{-5.2}$&$1.50^{+0.06}_{-0.08}$&$6670\pm220$&$1.10\pm0.10$&$1.20^{+0.08}_{-0.01}$&$0.060\pm0.020$&$-0.40\pm0.10$& [49]\\
EBLM-J2017+02Ab&$0.82$&$142.2^{+6.6}_{-6.7}$&$1.49^{+0.13}_{-0.10}$&$6161\pm124$&$1.11\pm0.07$&$1.20^{+0.08}_{-0.05}$&$0$ (fixed)&$-0.07\pm0.14$& [41]\\
KIC-1571511B&$14.02$&$148.1^{+0.5}_{-0.4}$&$1.74^{+0.00}_{-0.01}$&$6195\pm50$&$1.27^{+0.04}_{-0.03}$&$1.34\pm0.01$&$0.327\pm0.003$&$0.37\pm0.08$& [53]\\
WTS-19G-4-02069B&$2.44$&$149.8\pm6.3$&$1.69\pm0.06$&$3300\pm140$&$0.53\pm0.02$&$0.51\pm0.01$&$0$ (fixed)&-& [54]\\
\hline
\multicolumn{10}{l}{}
\end{tabular}
\label{tab:full_objs_lists_cont}
\end{table*}

\begin{table}
\caption{The derived parameters for \systemtnospace, produced by \textsc{Allesfitter}.}              
\label{tab:derived_parameters_full}      
\centering   
\begin{tabular}{l c}          
\hline
\hline
Parameter & Value\\ 
\hline 
\multicolumn{2}{c}{\textit{Derived parameters}} \\ 
\hline 
Host radius over semi-major axis b; $R_\star/a_\mathrm{b}$ & $0.01667\pm0.00020$  \\ 
Semi-major axis b over host radius; $a_\mathrm{b}/R_\star$ & $59.98\pm0.73$  \\ 
Companion radius b over semi-major axis b; $R_\mathrm{b}/a_\mathrm{b}$ & $0.001549\pm0.000032$  \\ 
Companion radius b; $R_\mathrm{b}$ ($\mathrm{R_{\oplus}}$) & $11.20\pm0.19$  \\ 
Companion radius b; $R_\mathrm{b}$ ($\mathrm{R_{jup}}$) & $0.999\pm0.017$  \\ 
Semi-major axis b; $a_\mathrm{b}$ ($\mathrm{R_{\odot}}$) & $66.3\pm1.1$  \\ 
Semi-major axis b; $a_\mathrm{b}$ (AU) & $0.3082\pm0.0050$  \\ 
Inclination b; $i_\mathrm{b}$ (deg) & $89.197\pm0.029$  \\ 
Eccentricity b; $e_\mathrm{b}$ & $0.77989\pm0.00049$  \\ 
Argument of periastron b; $w_\mathrm{b}$ (deg) & $214.482\pm0.087$  \\ 
Mass ratio b; $q_\mathrm{b}$ & $0.0698\pm0.0012$  \\ 
Companion mass b; $M_\mathrm{b}$ ($\mathrm{M_{\oplus}}$) & $23400\pm750$  \\ 
Companion mass b; $M_\mathrm{b}$ ($\mathrm{M_{jup}}$) & $73.6\pm2.4$  \\ 
Companion mass b; $M_\mathrm{b}$ ($\mathrm{M_{\odot}}$) & $0.0703\pm0.0023$  \\ 
Impact parameter b; $b_\mathrm{tra;b}$ & $0.590\pm0.015$  \\ 
Total transit duration b; $T_\mathrm{tot;b}$ (h) & $7.925\pm0.054$  \\ 
Full-transit duration b; $T_\mathrm{full;b}$ (h) & $5.935\pm0.061$  \\ 
Host density from orbit b; $\rho_\mathrm{\star;b}$ (cgs) & $1.121\pm0.041$  \\ 
Companion density b; $\rho_\mathrm{b}$ (cgs) & $91.6_{-5.6}^{+5.9}$  \\ 
Companion surface gravity b; $g_\mathrm{b}$ (cgs) & $182200\pm7500$  \\ 
Equilibrium temperature b; $T_\mathrm{eq;b}$ (K) & $464.2\pm7.2$  \\ 
Transit depth (undil.) b; $\delta_\mathrm{tr; undil; b; NGTS_2}$ (ppt) & $9.72_{-0.22}^{+0.26}$  \\ 
Transit depth (dil.) b; $\delta_\mathrm{tr; dil; b; NGTS_2}$ (ppt) & $9.72_{-0.22}^{+0.26}$  \\ 
Transit depth (undil.) b; $\delta_\mathrm{tr; undil; b; TESS_32}$ (ppt) & $9.61_{-0.36}^{+0.38}$  \\ 
Transit depth (dil.) b; $\delta_\mathrm{tr; dil; b; TESS_32}$ (ppt) & $8.72\pm0.19$  \\ 
Transit depth (undil.) b; $\delta_\mathrm{tr; undil; b; TESS_5}$ (ppt) & $9.59_{-0.38}^{+0.41}$  \\ 
Transit depth (dil.) b; $\delta_\mathrm{tr; dil; b; TESS_5}$ (ppt) & $8.94_{-0.20}^{+0.22}$  \\ 
Limb darkening; $u_\mathrm{1; NGTS_2}$ & $0.3756\pm0.0013$  \\ 
Limb darkening; $u_\mathrm{2; NGTS_2}$ & $0.4075\pm0.0014$  \\ 
Limb darkening; $u_\mathrm{1; TESS_32}$ & $0.3163\pm0.0011$  \\ 
Limb darkening; $u_\mathrm{2; TESS_32}$ & $0.3957\pm0.0012$  \\ 
Limb darkening; $u_\mathrm{1; TESS_5}$ & $0.3163\pm0.0011$  \\ 
Limb darkening; $u_\mathrm{2; TESS_5}$ & $0.3957\pm0.0012$  \\ 
Combined host density from all orbits; $rho_\mathrm{\star; combined}$ (cgs) & $1.121\pm0.041$ \\ 
\hline
\end{tabular}
\end{table}

\subsection{Corner Plots}\label{appendix:plots}

\begin{figure*}
    \centering
    \includegraphics[width=\linewidth, trim=0cm 0cm 0cm 2cm, clip]{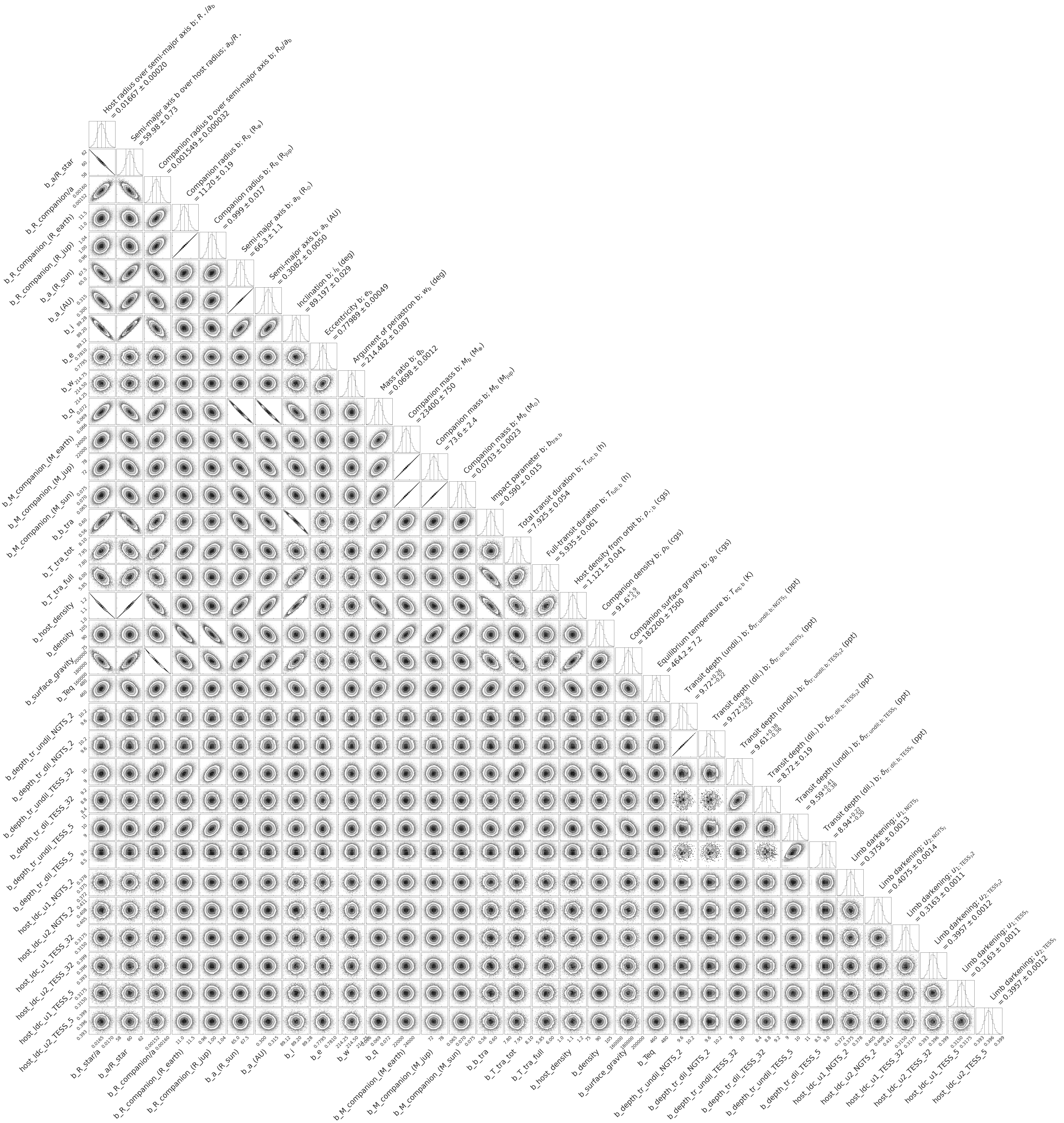}
    \caption{The corner plot for the derived values in the final fit.}
\label{fig:derived_corner}
\end{figure*}

\bsp	
\label{lastpage}
\end{document}